\documentclass[11pt,prd,aps,epsfig,floats,preprint,eqsecnum,nofootinbib]{revtex4}

\usepackage{color}
\usepackage[squaren]{SIunits}
\usepackage{amsmath}
\usepackage{amsfonts}
\usepackage{slashed}
\usepackage{graphicx}
\usepackage{subfigure}
\usepackage{rotating}

\usepackage[hypertex]{hyperref}
\usepackage{graphicx,amsmath,amssymb,epsfig,verbatim,mathrsfs,array,layout,textcomp,latexsym}
\usepackage{multirow}

\allowdisplaybreaks[1]

\renewcommand{\(}{\left(}
\renewcommand{\)}{\right)}
\renewcommand{\{}{\left\lbrace}
\renewcommand{\}}{\right\rbrace}
\renewcommand{\[}{\left\lbrack}
\renewcommand{\]}{\right\rbrack}
\renewcommand{\Re}[1]{\mathrm{Re}\!\{#1\}}

\newcommand{\dd}[1][{}]{\mathrm{d}^{#1}\!\!\;}

\newcommand{\refeq}[1]{Eq.~(\ref{eq:#1})}
\newcommand{\refeqs}[2]{Eqs.~(\ref{eq:#1})-(\ref{eq:#2})}
\newcommand{\reffig}[1]{Fig.~\ref{fig:#1}}
\newcommand{\refsec}[1]{Section \ref{sec:#1}}
\newcommand{\reftab}[1]{Table \ref{tab:#1}}

\def \re{\textrm{Re}}
\def \im{\textrm{Im}}

\newcommand{\alphae}{\alpha_\mathrm{e}}
\newcommand{\gfermi}{G_\mathrm{F}}
\newcommand{\GeV}{\,\mathrm{GeV}}
\newcommand{\MeV}{\,\mathrm{MeV}}

\newcommand{\wilson}[2][{}]{\mathcal{C}_{#2}^{\mathrm{#1}}}
\newcommand{\bra}[1]{\left\langle{#1}\right\vert}
\newcommand{\ket}[1]{\left\vert{#1}\right\rangle}

\def \azeL{{A_0^L}}
\def \azeR{{A_0^R}}
\def \apaL{{A_\|^L}}
\def \apaR{{A_\|^R}}
\def \apeL{{A_\bot^L}}
\def \apeR{{A_\bot^R}}

\def \thl {{\theta_l}}
\def \thK {{\theta_{K^*}}}

\begin{document}

\setlength{\parindent}{0pt}

\vspace{-0.2cm}
\hbox{CERN-PH-TH-2010-134}
\vspace{-0.2cm}
\hbox{DO-TH 10/08}

\title{The Benefits  of $\bar B \to \bar K^*  l^+ l^-$ Decays at Low Recoil}

\author{Christoph Bobeth}
\affiliation{IPHC, Universit\'e de Strasbourg, CNRS/IN2P3, F-67037, Strasbourg, France}
\author{Gudrun Hiller}
\affiliation{CERN, Theory Division, CH-1211 Geneva 23, Switzerland }
\affiliation{Institut f\"ur Physik, Technische Universit\"at Dortmund, D-44221 Dortmund, Germany}
\author{Danny van Dyk}
\affiliation{Institut f\"ur Physik, Technische Universit\"at Dortmund, D-44221 Dortmund, Germany}

\begin{abstract} 
Using the heavy quark effective theory framework put forward by Grinstein and
Pirjol we work out predictions for $ \bar B \to \bar K^* l^+ l^-$, $l=e,\mu$,
decays for a softly recoiling $\bar K^*$, {\it i.e.}, for large dilepton masses
$\sqrt{q^2}$ of the order of the $b$-quark mass $m_b$. We work to lowest order
in $\Lambda/Q$, where $Q=(m_b, \sqrt{q^2} )$ and include the next-to-leading
order corrections from the charm quark mass $m_c$ and the strong coupling at
${\cal{O}}(m_c^2/Q^2, \alpha_s)$.   The leading $\Lambda/m_b$ corrections are
parametrically suppressed. The improved Isgur-Wise form factor relations
correlate the $ \bar B \to \bar K^* l^+ l^-$ transversity amplitudes, which
simplifies the description of the various decay observables and provides
opportunities for the extraction of the electroweak short distance
couplings. We propose new angular observables which have very small hadronic uncertainties.
We exploit existing data on $ \bar B \to \bar K^* l^+ l^-$ distributions and show that the
low recoil region provides powerful additional information to the large recoil one.
We find disjoint best-fit solutions, which include the 
Standard Model, but also beyond-the-Standard Model ones. 
This ambiguity can be accessed with future precision measurements.

\end{abstract}

\maketitle

%
%
%--------+---------+---------+---------+---------+---------+---------+---------+
\section{Introduction}

The study of $b$-flavored mesons made possible our current understanding of
quark flavor violation in the Standard Model (SM) \cite{Cabibbo:1963yz}.  It is
an ongoing endeavour to map out the flavor sector at the electroweak scale and
beyond, and possibly thereby gaining insights on the origin of flavor.

In this effort, flavor changing neutral current-induced exclusive $B$ decays
into dileptons are important modes because of their sensitivity to physics
beyond the SM and their accessibility at current collider experiments and
possible future high luminosity facilities \cite{Anikeev:2001rk}.

We focus in this work on the semileptonic decays $\bar{B} \to \bar{K}^* l^+ l^-$
with $l=e, \mu$. Their branching ratios are measured at ${\cal{O}}(10^{- 7}
-10^{-6})$ \cite{Barberio:2008fa}, consistent with the SM \cite{Ali:1999mm}.
Beyond the rate, several observables can be obtained from the rare decays, in
particular when analyzed through $\bar{B} \to \bar{K}^* (\to \bar K \pi) l^+
l^-$ \cite{Kruger:1999xa}. The presence of multiple observables is advantageous
because they are, in general, complementary in their sensitivity to the
electroweak couplings, and they provide opportunities to control
uncertainties. This is even more important nowadays, as flavor physics data are
favoring the amount of fundamental flavor violation being at least not far away
from the one in the SM, and require a certain level of precision to be observed.

Recently, data have become available on $\bar{B} \to \bar{K}^* l^+ l^-$ decay
distributions in the dilepton invariant mass, $\sqrt{q^2}$, from the experiments
BaBar \cite{Aubert:2006vb, :2008ju}, Belle \cite{:2009zv} and CDF \cite{CDF2010}.
These experimental studies cover essentially the full kinematical dilepton mass
range, with the exception of the regions around $q^2 \sim m_{J/\psi}^2$ and $q^2
\sim m_{\psi^\prime}^2$. Here, cuts are employed to remove the overwhelming
background induced by $\bar{B} \to \bar{K}^{*} (\bar c c) \to \bar{K}^{*} l^+
l^-$ from the dominant charmonium resonances $(\bar c c) =J/\psi, \psi^\prime$.

Most theoretical works on $\bar{B} \to \bar{K}^{*} l^+ l^-$ decays over the past
years have focussed on the region of large recoil, that is, small $q^2 \lesssim
m_{J/\psi}^2$. However, at low recoil (large $q^2 \gtrsim m_{\psi\prime}^2$) dedicated studies 
are lacking with a similar QCD-footing as the ones at large
recoil, where QCD factorization (QCDF) applies \cite{Beneke:2001at, Beneke:2004dp}.  It
is the goal of this work to fill this gap and benefit from the incoming and
future physics data from the low recoil region as well.

We use the heavy quark effective theory (HQET) framework by Grinstein and Pirjol
\cite{Grinstein:2004vb}, which is applicable to the low recoil region, where
$\sqrt{q^2}$ is of the order of the mass of the $b$-quark, $m_b$, and the
emitted vector meson is soft in the $B$ mesons rest frame.
The original application was to extract the Cabibbo Kobayashi Maskawa (CKM) matrix element $V_{ub}$ by
relating the dilepton spectra of $\bar B \to \rho l \nu$ to those in $\bar{B}
\to \bar{K}^* l^+ l^-$ decays.  The framework has also been used previously to
study the implications of the sign of the forward-fackward asymmetry in $\bar B
\to \bar K^* l^+ l^-$ decays being determined SM-like for large $q^2$
\cite{Bobeth:2008ij}, see also \cite{HW} for relating $\bar{B} \to \bar{K} l^+
l^-$ to $\bar{B} \to \bar{K} \nu \bar \nu$ decays.  
 Here, we work out and
analyze in detail distributions of $\bar B \to \bar K^* l^+ l^-$ decays in this
low recoil framework and give predictions within the SM and beyond.

The description of $\bar B \to \bar K^* l^+ l^-$ decays at low recoil is based
on two ingredients: the improved Isgur-Wise form factor relations
\cite{Grinstein:2002cz,Grinstein:2004vb}, going beyond the original ones
\cite{Isgur:1990kf}, and an operator product expansion (OPE) in $1/Q$, where
$Q=(m_b, \sqrt{q^2})$ \cite{Grinstein:2004vb}. The latter allows to include the
contributions from quark loops, most notably charm loops in a model-independent
way. Both ingredients are first principle effective field theory tools and allow
to obtain the $\bar B \to \bar K^* l^+ l^-$ matrix element in a systematic
expansion in the strong coupling and in power corrections suppressed by the
heavy quark mass.
The implementation of continuum and resonance $\bar c c$
effects  from $e^+ e^- \to hadrons$ data
\cite{Kruger:1996cv} suggests no large duality violation at least above the
$\psi^\prime$, supporting the aforementioned OPE.

We work to lowest order in $\Lambda/m_b$, however, the actual leading power
corrections to the decay amplitudes arise only at order $\alpha_s \Lambda/m_b$
or with other parametric suppression factors, and amount only to a few percent.

The plan of the paper is as follows: In Section \ref{sec:general} we give the
electroweak Hamiltonian responsible for $b \to s l^+ l^-$ processes and review
the observables in $\bar{B} \to \bar{K}^* l^+ l^-$ decays.  The low recoil
framework is summarized in Section \ref{sec:loreco}, where the $\bar{B} \to
\bar{K}^* l^+ l^-$ transversity amplitudes and observables are computed and
correlations are pointed out. SM predictions and the comparison with the data
are given in Section \ref{sec:SMdata}. We conclude in Section \ref{sec:con}. In
several appendices we give formulae and detailed input for our analysis.

%
%
%--------+---------+---------+---------+---------+---------+---------+---------+
\section{Generalities \label{sec:general}}

We define the short distance couplings entering $b \to s l^+ l^-$ decays in
Section \ref{sec:quark} and introduce in Section \ref{sec:observables} the
observables in $\bar B \to \bar K^* l^+ l^-$ decays, where the former can be
tested.

%
%--------+---------+---------+---------+---------+---------+---------+---------+
\subsection{Quark level \label{sec:quark}}
For the description of processes induced by $b \to s l^+ l^-$ we use an
effective $\Delta B=1$ electroweak Hamiltonian
\begin{align}
  {\cal{H}}_{\rm{eff}} & = 
  - \frac{4 \gfermi}{\sqrt{2}}  V_{tb}
V^*_{ts}  \sum_i 
    \wilson{i}(\mu) \mathcal{O}_i(\mu) 
  + {\rm h.c.},
\end{align}
which consists of the higher dimensional  operators $\mathcal{O}_i$ and their
respective Wilson coefficients $\wilson{i}$. Here, $\mu$ denotes the
renormalization scale, $\gfermi$ is Fermi's constant and $V_{tb}
V^*_{ts}$ collects the leading flavor factors of the SM encoded in the CKM
matrix elements $V_{ij}$.  We neglect subleading contributions of the order
$V_{ub} V_{us}^*$, hence, there is no CP violation in the SM in the decay
amplitudes. We also set the strange quark mass to zero.

For the decays $b \to s l^+ l^-$ the electromagnetic dipole ($\mathcal{O}_7$)
and semileptonic four-fermion ($\mathcal{O}_{9,10}$) operators are the most
relevant:
\begin{align} 
  \mathcal{O}_7 & = 
    \frac{e}{(4\pi)^2} m_b \[\bar{s} \sigma^{\mu\nu} P_{R} b\] F_{\mu\nu} ,
& \mathcal{O}_8 & = 
    \frac{g_s}{(4\pi)^2} m_b \[\bar{s} \sigma^{\mu\nu} P_{R} T^a b\] G^a_{\mu\nu} ,\        
\nonumber \\
  \mathcal{O}_9 & = 
    \frac{e^2}{(4\pi)^2} \[\bar{s} \gamma_\mu P_{L} b\] \[\bar{l} \gamma^\mu l\] ,
& \mathcal{O}_{10} & = 
    \frac{e^2}{(4\pi)^2} \[\bar{s} \gamma_\mu P_{L} b\] \[\bar{l} \gamma^\mu \gamma_5 l\] ,
  \label{eq:smbasis}
\end{align}
where $P_{L,R}$ denote chiral projectors, $m_b$ is the $\overline{\rm MS}$ mass
of the $b$-quark and $F_{\mu \nu}(G^a_{\mu \nu}) $ is the field strength tensor
of the photon (gluons $a=1, ...,8$).  The contributions from the gluonic dipole
operator $\mathcal{O}_8$ enter the semileptonic decay amplitude at higher order
in the strong coupling $g_s$, and have a significantly reduced sensitivity to
New Physics as compared to those from $\mathcal{O}_{7,9,10}$.  For the current-current
and QCD-penguin
operators $\mathcal{O}_{1 \ldots 6}$  we use the
definitions of Ref.~\cite{Chetyrkin:1996vx}.  We call the set of operators
\refeq{smbasis} plus the four-quark operators $\mathcal{O}_{1 \ldots 6}$ the
SM basis, and stay in this work within this basis.

The goal of this work is to extract from $b$-physics data the coefficients
$\wilson{7,9,10}$ and test them against their respective SM predictions. All
other Wilson coefficients are fixed to their respective SM values. We restrict
ourselves to real-valued  Wilson coefficients, hence allow for no CP violation beyond
the SM.  We made this choice because existing CP data on the $b \to s l^+ l^-$
transitions \cite{Barberio:2008fa}, which are consistent with our assumption,
are currently quite limited, have rather large uncertainties, and the inclusion
of phases doubles the number of parameters in the fit. We hope to come back to
this in the future.

In the following we understand all Wilson coefficients being evaluated at the
scale of the $b$-quark mass.  In the SM at next-to-leading order their values
are approximately, for $\mu=m_b$,
\begin{align}
  \mathcal{C}_7^{\rm SM} & = -0.3 , &
  \mathcal{C}_9^{\rm SM} & = 4.2 ,  &
  \mathcal{C}_{10}^{\rm SM} & = -4.2. 
\end{align}
The coefficient of $\mathcal{O}_{7}$ is suppressed with respect to the ones of
$\mathcal{O}_{9,10}$, a feature that holds in many extensions of the SM as well,
and is also respected by the data. This hierarchy in coupling strengths is
beneficial for controlling theoretical uncertainties, see Section
\ref{sec:loreco}.

We neglect lepton flavor non-universal effects, hence, the couplings to $l=e$
and $l=\mu$ are considered to be equal. For recent works exploiting the
possibility that New Physics affects the final state electron and muon pairs
differently, see, {\it e.g.,} \cite{Bobeth:2007dw}. Since the decays $b \to s
\tau^+ \tau^-$ are experimentally difficult and have not been seen so far, we do
not consider taus and can neglect the lepton masses.

%
%--------+---------+---------+---------+---------+---------+---------+---------+
\subsection{The $\bar B \to \bar K^* l^+ l^-$ observables \label{sec:observables}}

Angular analysis offers the maximal information which is accessible from the
decay via $\bar{B} \to \bar{K}^* (\to \bar{K} \pi) l^+ l^-$. 
%(We use $\bar{B} \equiv (b\bar{q})$, $q=u,d$.)  
For an on-shell $\bar{K}^*$ the differential
decay width can be written as \cite{Kruger:1999xa,Kruger:2005ep}
\begin{align}
  \label{eq:decaydistribution}
  \frac{\dd[4] \Gamma}{\dd q^2 \dd\cos\theta_l \dd\cos\theta_{K^*} \dd\phi}
  & = \frac{3}{8\pi} J(q^2, \cos\theta_l, \cos\theta_{K^*}, \phi),
\end{align}
where the lepton spins have been summed over.  Here, $q^2$ is the dilepton
invariant mass squared, that is, $q^\mu$ is the sum of $p_{l^+}^\mu$ and
$p_{l^-}^\mu$, the four momenta of the positively and negatively charged lepton,
respectively.  Furthermore, $\thl$ is defined as the angle between the
negatively charged lepton and the $\bar{B}$ in the dilepton center of mass
system (c.m.s.) and $\thK$ is the angle between the Kaon and the $\bar{B}$ in
the $(K^-\pi^+)$ c.m.s.. We denote by $\mathbf{p}_i$ the three momentum vector
of particle $i$ in the $\bar{B}$ rest frame. Then, $\phi$ is given by the angle
between $\mathbf{p}_{K^-} \times \mathbf{p}_{\pi^+}$ and $\mathbf{p}_{l^-}
\times \mathbf{p}_{l^+}$, {\it i.e.}, the angle between the normals of the
$(K^-\pi^+)$ and $(l^-l^+)$ planes.

The full kinematically accessible phase space is bounded by
\begin{align}
  4 m_l^2 & \leqslant q^2\leqslant (m_B - m_{K^*})^2, &
  -1 & \leqslant \cos\theta_l \leqslant 1, &
  -1 & \leqslant \cos\theta_{K^*} \leqslant 1, &
   0 & \leqslant \phi     \leqslant 2 \pi, 
\end{align}
where $m_l, m_B$ and $m_{K^*}$ denote the mass of the lepton, $B$ meson and the
$K^*$, respectively.

The dependence of the decay distribution \refeq{decaydistribution} on the angles
$\theta_l,\, \theta_{K^*}$ and $\phi$ can be made explicit as
\begin{align}
  J(q^2, \theta_l, \theta_{K^*}, \phi)& = J_1^s \sin^2\theta_{K^*} + J_1^c \cos^2\theta_{K^*}
      + (J_2^s \sin^2\theta_{K^*} + J_2^c \cos^2\theta_{K^*}) \cos 2\theta_l
\nonumber \\
    & + J_3 \sin^2\theta_{K^*} \sin^2\theta_l \cos 2\phi
      + J_4 \sin 2\theta_{K^*} \sin 2\theta_l \cos\phi
      + J_5 \sin 2\theta_{K^*} \sin\theta_l \cos\phi
\nonumber \\
    & + J_6 \sin^2\theta_{K^*} \cos\theta_l
      + J_7 \sin 2\theta_{K^*} \sin\theta_l \sin\phi
\nonumber \\
    & + J_8 \sin 2\theta_{K^*} \sin 2\theta_l \sin\phi
      + J_9 \sin^2\theta_{K^*} \sin^2\theta_l \sin 2\phi,
\end{align}
where the angular coefficients $J_i^{(a)} =J_i^{(a)}(q^2)$ for $i = 1, \ldots,
9$ and $a = s,c$ are functions of the dilepton mass. We
suppress in the following the $q^2$-dependence also in expressions derived from the
$J_i^{(a)}$.  The latter can be written in terms of the transversity amplitudes
$A_{\perp,\parallel,0}$, see Appendix \ref{sec:JifromAi}.  The fourth amplitude
$A_t$ does not contribute in the limit $m_l=0$. The transversity amplitudes at
low recoil are given in the next section. The ones at large recoil can be seen,
for example, in Ref.~\cite{Bobeth:2008ij}.

The angular coefficients $J_i^{(a)}$, or their normalized variants $J_i/(\dd
\Gamma / \dd q^2)$ or $J_i/J_j$, are observables which can be extracted from an angular
analysis.  This method allows to test the SM and probe a multitude of different
couplings \cite{Bobeth:2008ij,Kruger:2005ep,Egede:2008uy,Altmannshofer:2008dz,Egede:2010zc}.  We focus
first on rather simple observables, which can be extracted without performing a
statistics intense full angular analysis. Afterwards, we point out opportunities
of measuring the angular distribution.

Data on $\bar B \to \bar K^* l^+ l^-$ decays already exists from BaBar
\cite{Aubert:2006vb,:2008ju}, Belle \cite{:2009zv} and CDF \cite{CDF2010} for
the differential decay width
$\dd \Gamma / \dd q^2$, the forward-backward asymmetry $A_{\rm FB}$ and the
fraction of longitudinal polarized $K^*$'s, $F_{\rm L}$. They are written as
\begin{align} 
  \label{eq:dGam:def}
  \frac{\dd \Gamma}{\dd q^2} & = 
  2 J_1^s+J_1^c-\frac{2 J_2^s+J_2^c}{3} = 
  |A_0^L|^2 + |A_\perp^L|^2 + |A_\parallel^L|^2 + (L \leftrightarrow R),
\\  
  A_{ \rm FB} &  = 
  \left[\int_{0}^1 - \int_{-1}^0 \right] \dd \! \cos\theta_l\, 
  \frac{\dd^2\Gamma}{\dd q^2 \, \dd\! \cos\theta_l} \Bigg/\frac{\dd \Gamma}{\dd q^2} = 
  \frac{J_6}{\dd \Gamma / \dd q^2}, 
  \label{eq:Afb} 
\\
  F_{\rm L} & =
  \frac{|A_0^L|^2 + |A_0^R|^2}{\dd \Gamma / \dd q^2},  
  \label{eq:FL}
\end{align}
and are all distributions in the dilepton mass.

The experimental data on the $q^2$-distributions
\cite{:2009zv,Aubert:2006vb,:2008ju,CDF2010} are currently available in
$q^2$-bins, {\it i.e.}, the decay rate is given as a list of rates $\langle \dd
\Gamma/\dd q^2 \rangle_k$, where we denote by $\langle .. \rangle_k$ the $\dd
q^2$-integration over the $k$-th bin.  Normalized quantities such as the
forward-backward asymmetry are then delivered as $\langle J_6 \rangle_k/\langle
\dd \Gamma/\dd q^2 \rangle_k$, and likewise as $\langle |A_0^L|^2 + |A_0^R|^2
\rangle_k/\langle \dd \Gamma/\dd q^2 \rangle_k$ for the longitudinal
polarization fraction. The binned distributions equal our definitions
Eqs.~(\ref{eq:Afb}) and (\ref{eq:FL}) for flat distributions or infinitely small
bin size.

Note that the $J_{5,6,8,9}$, and hence $A_{\rm FB}$ are CP-odd observables, which
vanish in an untagged equally mixed sample of $\bar B$ and $B$ decays in the
absence of CP violation \cite{Bobeth:2008ij}.

We also consider the transverse asymmetries $A_T^{(2)}$
\cite{Kruger:2005ep} and $ A_T^{(3,4)}$ \cite{Egede:2008uy}, given as
\begin{align}
  A_T^{(2)} & =
  \frac{|A_\perp^L|^2 + |A_\perp^R|^2-|A_\parallel^L|^2-|A_\parallel^R|^2}
       {|A_\perp^L|^2 + |A_\perp^R|^2+|A_\parallel^L|^2+|A_\parallel^R|^2}
 = \frac{1}{2} \frac{J_3}{J_2^s} , 
\\
  A_T^{(3)} & = 
  \frac{\vert A_0^L A_\parallel^{L*} + A_0^{R*} A_\parallel^R\vert}
       {\sqrt{\big(\vert A_0^L\vert^2 + \vert A_0^R\vert^2\big)
              \big(\vert A_\perp^L\vert^2 + \vert A_\perp^R\vert^2\big)}}
 = \sqrt{\frac{4 J_4^2 + \beta_l^2 J_7^2}{- 2 J_2^c (2 J_2^s + J_3)}} ,
\\
  \label{eq:AT4:def}
  A_T^{(4)} & = 
  \frac{\vert A_0^L A_\perp^{L *} - A_0^{R*} A_\perp^R\vert}
       {\vert A_0^{L*} A_\parallel^L + A_0^R A_\parallel^{R*}\vert}
 = \sqrt{ \frac{\beta_l^2 J_5^2 + 4 J_8^2}{4 J_4^2 + \beta_l^2 J_7^2}} ,
\end{align}
which have not been measured yet. The factor $\beta_l$ is given in
Appendix \ref{sec:JifromAi}. Here we keep the lepton mass dependence
for generality but discard it later on when discussing the low recoil region
where $m_l$ is entirely neglibile.

We propose the following new transversity observables for the region of low recoil (high $q^2$)
\begin{align}
  H_T^{(1)} & =  
    \frac{{\rm Re}( A_0^L A_\parallel^{L*} + A_0^{R*} A_\parallel^R)}
     {\sqrt{\big(\vert A_0^L\vert^2 + \vert A_0^R\vert^2\big)
            \big(\vert A_\parallel^L\vert^2 + \vert A_\parallel^R\vert^2\big)}}
  = \frac{\sqrt{2} J_4}{\sqrt{- J_2^c \left(2 J_2^s - J_3\right)}} ,
  \label{eq:def:HT1}
\\
  H_T^{(2)} & = 
    \frac{{\rm Re}( A_0^L A_\perp^{L*} - A_0^{R*} A_\perp^R)}
     {\sqrt{\big(\vert A_0^L\vert^2 + \vert A_0^R\vert^2\big)
            \big(\vert A_\perp^L\vert^2 + \vert A_\perp^R\vert^2\big)}}
  = \frac{\beta_l J_5}{\sqrt{-2 J_2^c \left(2 J_2^s + J_3\right)}} ,
  \label{eq:def:HT2}
\\
  H_T^{(3)} & =
    \frac{{\rm Re}( A_\parallel^L A_\perp^{L*} - A_\parallel^{R*} A_\perp^R)}
     {\sqrt{\big(\vert A_\parallel^L\vert^2 + \vert A_\parallel^R\vert^2\big)
            \big(\vert A_\perp^L\vert^2 + \vert A_\perp^R\vert^2\big)}}
  = \frac{\beta_l J_6}{2 \sqrt{(2 J_2^s)^2 - J_3^2}}.
  \label{eq:def:HT3}
\end{align} 
As will become clear in Section \ref{sec:loreco}, see also Appendix \ref{sec:Ui},
the $H_T^{(i)}$ are designed to have very small hadronic uncertainties at low recoil.
While both $H_T^{(3)}$ and  $A_{\rm FB}$ depend on $J_6$ and probe similar short distance physics, the former has a significantly smaller theoretical uncertainty than the latter.
Note also that the
numerator $J_5$ of $H_T^{(2)}$ is related to the observable $S_5$ which has good
 prospects to be measured with early LHC$b$ data of 2 fb${}^{-1}$ at least in the large recoil region \cite{Bharucha:2010bb}.
 
Different possibilities to extract the $J_i$ from
single differential distributions as well have been outlined in \cite{Bobeth:2008ij}.

%
%--------+---------+---------+---------+---------+---------+---------+---------+
\section{$\bar B \to \bar K^* l^+ l^-$  at low recoil \label{sec:loreco}}

We start in Section \ref{sec:ope} with the model-independent description of the
exclusive heavy-to-light decays in the low recoil region following Grinstein and
Pirjol \cite{Grinstein:2004vb,Grinstein:2002cz}. After calculating and
investigating the $\bar B \to \bar K^* l^+ l^-$ transversity amplitudes in
Section \ref{sec:transA}, we work out predictions for and correlations between
the $\bar B \to \bar K^* l^+ l^-$ observables at low recoil in Section
\ref{sec:predictions}.  A numerical study within the SM is given in Section
\ref{sec:smstatus}.
 
%
%--------+---------+---------+---------+---------+---------+---------+---------+
\subsection{The  model-independent framework  \label{sec:ope}}
 
The description of $\bar B \to \bar K^* l^+ l^-$ decays at low recoil, where
$q^2 \sim {\cal{O}}(m_b^2)$, is based on the improved form factor relations in
this region and an OPE in $1/Q$ \cite{Grinstein:2004vb,Grinstein:2002cz}. The latter keeps
the non-perturbative contributions from 4-quark operators $(\bar s
b) (\bar q q)$ under control by expanding in $m_q^2/Q^2$.  This is most important for 
charm quarks,
since their operators can enter with no suppression from small Wilson
coefficients nor CKM matrix elements.
 
Following \cite{Grinstein:2004vb} we briefly sketch the derivation of the
improved Isgur-Wise form factor relations to leading order in $1/m_b$ between the
vector and the tensor current. The starting point is the QCD operator identity
(for $m_s=0$)
\begin{align} 
  \label{eq:opID}
  i \partial^\nu (\bar s i \sigma_{\mu \nu} b) =
   - m_b \bar s \gamma_\mu b +i \partial_\mu (\bar s b) 
   - 2 \bar s i \!\stackrel{\leftarrow}{D}_{\mu} \! b.
\end{align}
After taking the matrix element of \refeq{opID} using the form factors given in
Appendix \ref{sec:ChoiceFF} one arrives at an exact relation between the form
factors $T_1$ and $V$ and the matrix element of the current $\bar s i
\!\stackrel{\leftarrow}{D}_{\mu} \! b $. The latter can be expanded in $1/m_b$
through matching onto the HQET currents with the heavy quark field $h_v$:
\begin{align} 
  \label{eq:newmat}
  \bar s i \!\stackrel{\leftarrow}{D}_{\mu} \! b & =
  D_0^{(v)} (\mu) m_b  \bar s \gamma_\mu h_v +
  D_1^{(v)} (\mu) m_b  v_\mu \bar s h_v + \dots .
\end{align}
We further need 
\begin{align}
  \bar s \gamma_\mu b & = 
  C_0^{(v)} (\mu) \bar s \gamma_\mu h_v + C_1^{(v)}(\mu)  v_\mu \bar s h_v + \dots,
\\
  \bar s  b & = 
  C_0^{(s)} (\mu) \bar s  h_v  + \dots ,
\end{align}
to express the HQET currents in \refeq{newmat} through quark currents. The
ellipses denote power suppressed contributions. The Wilson coefficients
$C_i^{(x)}$ and $D_i^{(x)}$ are calculable and known in a perturbative expansion
in the strong coupling, see, {\it e.g.}, \cite{Manohar:2000dt,Grinstein:2004vb}.

Taking then the matrix element of  \refeq{newmat}  yields
\begin{align}
  \langle  K^* | \bar s i \!\stackrel{\leftarrow}{D}_{\mu} \! b | B \rangle & = 
    \frac{m_b D^{(v)}_0 (\mu)}{C_0^{(v)}(\mu) } 
    \langle K^*|  \bar s \gamma_\mu b |B \rangle + \ldots .
\end{align}
After working out the corresponding formulae involving the axial currents, the
improved Isgur-Wise relations to leading order in $1/m_b$ including radiative
corrections are obtained as
\begin{align} 
  \label{eq:ffrelations}
    T_1(q^2)
        & = \kappa V(q^2),
&   T_2(q^2)
        & = \kappa A_1(q^2),
&   T_3(q^2)
        & = \kappa A_2(q^2) \frac{m_B^2}{q^2},
\end{align}
where
\begin{align}
  \label{eq:ffrelationsLO}
  \kappa = \(1 + \frac{2D_0^{(v)}(\mu)}{C_0^{(v)}(\mu)}\) \frac{m_b(\mu)}{m_B} .
\end{align}
Here, subleading terms of the order $m_{K^*}/m_B$, $\Lambda/m_B$ are dropped and a
naively anticommuting $\gamma_5$ matrix is used. The latter allows to
relate the HQET Wilson coefficients of currents without a $\gamma_5$ matrix to
those containing one by replacing $\bar s$ with $\bar s (-\gamma_5)$ in the
matching equations.
We also suppress the renormalization scale dependence of the penguin form
factors $T_i$ and of the coefficient $\kappa$. It reads, up to corrections of
${\cal{O}}(\alpha_s^2)$,
\begin{align}
  \kappa & =
  1 - 2\, \frac{\alpha_s}{3\pi} \ln \left(\frac{ \mu}{m_b}\right) .
\end{align}
The relations \refeq{ffrelations} are consistent with  the ones derived in
\cite{Grinstein:2004vb} at lowest order in $1/m_b$ after changing to the
Isgur-Wise form factor basis \cite{Isgur:1990kf}.

The inclusion of the 4-quark and gluon dipole operators leads to the effective
couplings, $\wilson[eff]{7,9}$ \cite{Grinstein:2004vb}. They read
\begin{align}  
  \label{eq:c9effGP}
  \wilson[eff]{9} & = 
    \wilson{9} + 
    h(0, q^2) \[ \frac{4}{3}\, \wilson{1} + \wilson{2} + \frac{11}{2}\, \wilson{3}
      - \frac{2}{3}\, \wilson{4} + 52\, \wilson{5} - \frac{32}{3}\, \wilson{6}\] 
\\
  & - \frac{1}{2}\, h(m_b, q^2) \[ 7\, \wilson{3} + \frac{4}{3}\, \wilson{4} + 76\, \wilson{5}
      + \frac{64}{3}\, \wilson{6} \]
    + \frac{4}{3} \[ \wilson{3} + \frac{16}{3}\, \wilson{5} + \frac{16}{9}\, \wilson{6} \]
\nonumber\\
  & + \frac{\alpha_s}{4 \pi} \[ \wilson{1} \left(B(q^2) + 4\, C(q^2)\right) 
      - 3\, \wilson{2}\left(2\, B(q^2) - C(q^2)\right) - \wilson{8}F_8^{(9)}(q^2) \]  
\nonumber\\ 
  & + 8 {\frac{m_c^2}{q^2}  \[\frac{4}{9}\,\wilson{1} 
      + \frac{1}{3}\,\wilson{2} + 2\,\wilson{3} + 20\,\wilson{5} \] },  
\nonumber \\
  \label{eq:c7effGP}
  \wilson[eff]{7} & = 
  \wilson{7} - 
  \frac{1}{3} \[ \wilson{3} + \frac{4}{3}\,\wilson{4} + 20\,\wilson{5} 
    + \frac{80}{3}\wilson{6} \]
  + \frac{\alpha_s}{4 \pi} \[ \left(\wilson{1} - 6\,\wilson{2}\right) A(q^2) 
    - \wilson{8} F_8^{(7)}(q^2)\] ,
\end{align}
and we recall that we use the 4-quark operators ${\cal{O}}_{1 ...6}$ as defined
in \cite{Chetyrkin:1996vx}.  The functions $A,B,C$ and $F_8^{(7)}, F_8^{(9)}$
can be seen in \cite{Seidel:2004jh} and \cite{Beneke:2001at}, respectively.\footnote{Note that in
\cite{Seidel:2004jh} a different sign convention has been used than in the
previous works \cite{Asatrian:2001de}.} The lowest order charm
loop function is given as
\begin{eqnarray} 
  \label{eq:charmloop}
  h(0,q^2) = \frac{8}{27} 
  + \frac{4}{9} \left(\ln \frac{\mu^2}{q^2} + i\pi \right) ,
\end{eqnarray}
which is simply  the perturbative quark loop function for massless quarks.  The
$m_c^2/Q^2$ corrections are given by the last line of \refeq{c9effGP}.
Loops with $b$ quarks stemming from penguin operators are taken into account by
the function
\begin{eqnarray} \label{eq:charmloopFull}
h(m_b,q^2) = \frac{4}{9}\(\ln \frac{\mu^2}{m_b^2} +\frac{2}{3} + z\)
             -\frac{4}{9}(2 + z)\sqrt{z - 1}
                \arctan \frac{1}{\sqrt{z - 1}} , ~~z = \frac{4m_b^2}{q^2} .
                \end{eqnarray}
We stress that the effective coefficients
Eqs.~(\ref{eq:c9effGP})-(\ref{eq:c7effGP}) are different from the ones
used in  the low $q^2$ region given in \cite{Beneke:2001at}.

The product $m_b\, \kappa\, \wilson[eff]{7}$ is independent of the
renormalization scale \cite{Grinstein:2004vb}. As we will see in the next
section, this is important because contributions from $\wilson[eff]{7}$ enter
the $\bar B \to \bar K^* l^+ l^-$ amplitudes in exactly this combination. The
$\mu$-dependence of $\wilson[eff]{9}$ is very small and induced at the order
$\alpha_s^2\, \wilson{1,2}$ and $\alpha_s\, \wilson{3,..6}$.

The heavy quark matrix elements $\langle K^* | \bar s i
\!\stackrel{\leftarrow}{D}_{\mu} \!  (\gamma_5) h_v | B \rangle$ 
are the only new hadronic input required at order $\Lambda/m_b$ for both the
form factor relations and the matrix elements related to the electromagnetic
current, $\wilson[eff]{7,9}$  \cite{Grinstein:2004vb}.
However, we refrain from including these explicit  $\Lambda/m_b$ corrections. Firstly, 
the requisite additional matrix elements are currently only known from constituent quark model calculations \cite{Grinstein:2002cz,GP0209} bringing in sizable uncertainties. 
More importantly,  the leading power corrections to the form factor relations are parametrically suppressed,  see \refsec{transA}. Note that the ones to the OPE arise only at ${\cal{O}}(\alpha_s
\Lambda/m_b, m_c^4/Q^4)$. Hence, the power corrections have a reduced impact on the decay observables. Quantitative estimates are given in \refsec{smstatus}.

Note that explicit spectator effects are power suppressed and absent to the
order we are working. They only appear indirectly in the form factors, lifetime
and meson masses. Hence, the formulae can be used for charged and neutral $ \bar
B \to \bar K^* l^+ l^-$ decays, and $\bar B_s \to \phi l^+ l^-$ decays after the
necessary replacements.

%--------+---------+---------+---------+---------+---------+---------+---------+
\subsection{The transversity amplitudes  \label{sec:transA}}

Application of the form factor relations in \refeq{ffrelations} and using the effective
coefficients Eqs.~(\ref{eq:c9effGP})-(\ref{eq:c7effGP}) yields the low recoil
transversity amplitudes to leading order in $1/m_b$ as
\begin{align}
  \label{eq:Aperp}
  A_\perp^{L,R} & = 
  +i \{(\wilson[eff]{9} \mp \wilson{10}) + 
      \kappa\frac{2 \hat{m}_b}{\hat{s}}\, \wilson[eff]{7}\}
      f_\perp ,
\\[1mm]
  A_\parallel^{L,R} & =
    -i \{(\wilson[eff]{9} \mp \wilson{10}) + 
      \kappa\frac{2 \hat{m}_b}{\hat{s}}\, \wilson[eff]{7}\} f_\parallel,
      \\
  \label{eq:A0}
  A_0^{L,R} & =
  -i \{\(\wilson[eff]{9} \mp \wilson{10}\) + 
    \kappa \frac{2 \hat{m}_b}{\hat{s}}\, \wilson[eff]{7}\} f_0 ,
    \end{align}
where the form factors enter as
\begin{align}
  f_\perp & =  N m_B \frac{\sqrt{2 \hat{\lambda}}}{1 + \hat m_{K^*}} V, &
  f_\parallel & =  N m_B \sqrt{2}\, (1 + \hat m_{K^*})\, A_1, 
  \nonumber
\end{align}
\begin{align}
  f_0 & = N m_B 
    \frac{(1 - \hat{s} - \hat m_{K^*}^2) (1 + \hat m_{K^*})^2 A_1 - \hat{\lambda}\, A_2}
    {2\, \hat m_{K^*} (1 + \hat m_{K^*}) \sqrt{\hat{s}}} ,
\end{align}
and the normalization factor reads
\begin{align}
  N & = 
    \sqrt{\frac{\gfermi^2\, \alphae^2\, |\lambda_t|^2\, m_B\, \hat s 
          \sqrt{ \hat \lambda}}{3 \cdot 2^{10}\, \pi^5}} .
\end{align}
Here, we switched to the dimensionless variables $\hat s =q^2/m_B^2$, $\hat m_i
=m_i/m_B$ and $\hat \lambda =1+ \hat s^2 + \hat m_{K^*}^4 - 2\, (\hat s + \hat s
\hat m_{K^*}^2 + \hat m_{K^*}^2)$.  We also suppressed for brevity the
dependence on the momentum transfer in the form factors and the effective
coefficients. We further neglected subleading terms of order $m_{K^*}/m_B$ in
the $\wilson[eff]{7}$-term only.

Interestingly, within our framework (SM basis, lowest order in $\Lambda/m_b$)
the transversity amplitudes Eqs.~(\ref{eq:Aperp})-(\ref{eq:A0}) depend in exactly the
same way on the short distance
coefficients. Consequently, only two independent combinations of Wilson
coefficients can be probed, related to $|A^L_i|^2 \pm |A^R_i|^2$, since $A^L$
and $A^R$ do not interfere for massless leptons, see Appendix
\ref{sec:JifromAi}. The independent combinations can be defined as
\begin{align} 
  \rho_1 & \equiv 
  \left|\wilson[eff]{9} + \kappa \frac{2\hat{m}_b}{\hat{s}}\wilson[eff]{7}\right|^2 
   + \left|\wilson{10}\right|^2 , 
\\
  \rho_2 & \equiv 
  \Re{\(\wilson[eff]{9} + \kappa \frac{2\hat{m}_b}{\hat{s}} \wilson[eff]{7}\) \wilson[*]{10}} .  
\end{align}
$\rho_1$ and $\rho_2$ are largely $\mu$-scale independent. The dominant
dependence on the dilepton mass in $\rho_{1,2}$ stems from the $1/\hat s$-factor
accompanying $\wilson[eff]{7}$.
The short distance parameter $\rho_1$  equals up to $\Lambda/m_b$ corrections
the parameter $N_{\rm eff}$ introduced in Ref.~\cite{Grinstein:2004vb}.
 
The relation between all three transversity amplitudes makes
the low recoil region overconstrained and very predictive. 
We work out the corresponding implications in Section \ref{sec:predictions}.
Note that in the
large recoil region two amplitudes are related as
$A_\parallel^X =-A^X_\perp$ by helicity conservation up to corrections in
$1/E_{K^*}$ in the SM basis \cite{Burdman:2000ku}.  

The leading power corrections of the OPE arise at ${\cal{O}}(\alpha_s
\Lambda/m_b, m_c^4/Q^4) \sim $ few percent.  The $\Lambda/m_b$ corrections to
the amplitudes from the form factor relations  are parametrically suppressed as
well, by small dipole coefficients, such that we can estimate the leading power
correction from the form factor relations to the decay amplitudes as order $( 2
\wilson[eff]{7}/ \wilson[eff]{9} ) \Lambda/m_b$. So in general, the dominant
power corrections to the transversity amplitudes are of the order few percent.

We simulate the effect of the $1/m_b$ corrections by dimensional analysis when
estimating theoretical uncertainties in Section \ref{sec:smstatus}.  

%
%--------+---------+---------+---------+---------+---------+---------+---------+
\subsection{Observables and predictions \label{sec:predictions}}

We begin with low recoil predictions of some basic distributions.  At leading
order they can be written in terms of the transversity amplitudes
$A_{\perp,\parallel,0}$ given in Eqs.~(\ref{eq:Aperp})-(\ref{eq:A0}) as:
\begin{align} \label{eq:dG:HQET}
  \frac{\dd \Gamma}{\dd q^2} & = 
  2\, \rho_1 \times (f_0^2 + f_\perp^2 + f_\parallel^2),
\\
  A_{\rm FB} & =
  3\, \frac{\rho_2}{\rho_1} \times \frac{f_\perp f_\parallel}
  {(f_0^2 + f_\perp^2 + f_\parallel^2),}
\\
  F_{\rm L} & = 
  \frac{f_0^2}{f_0^2 + f_\perp^2 + f_\parallel^2},
\end{align}
and            
\begin{align}                       
  A_T^{(2)} & = \frac{f_\perp^2 - f_\parallel^2}{f_\perp^2 + f_\parallel^2} , & 
  A_T^{(3)} & = \frac{f_\parallel}{f_\perp} , &
  A_T^{(4)} & = 2\, \frac{\rho_2}{\rho_1} \times \frac{f_\perp}{f_\parallel} .
\end{align}
The new high $q^2$ transversity observables read as
\begin{align}    \label{eq:Hi:HQET}                   
  H_T^{(1)} & = 1 , & 
  H_T^{(2)} & = H_T^{(3)} =2\, \frac{\rho_2}{\rho_1} .
  \end{align}
  All observables factorize into short distance coefficients $\rho_{1,2}$ and
  form factor ones $f_{0,\perp, \parallel}$.

We note the following:
\begin{itemize}
\item The only two independent combinations of Wilson coefficients, $\rho_1$ and
  $\rho_2$, enter the decay rate $\dd \Gamma/\dd q^2$ and the forward-backward
  asymmetry $A_{\rm FB}$, respectively.

\item The observables $F_{\rm L}$, $A_T^{(2,3)}$ and $H_T^{(1)}$ are independent
  of the Wilson coefficients.  Data on $F_{\rm L}$ and $A_T^{(2,3)}$ test the
  form factors. In particular, $A_T^{(2)}$ and $A_T^{(3)}$ each measure the
  ratio $A_1/V$, whereas $F_{\rm L}$ is in addition sensitive to $A_2$.  More
  observables designed to not depend on the short distance coefficients are
  given in Appendix \ref{sec:Ui},  see \refeq{ff-ratio-test}.
 
\item More generally, in the SM basis and to the order we are working, any
  observable in the decay $\bar{B} \to \bar{K}^* (\to \bar{K} \pi) l^+ l^-$ is 
  correlated with $\dd \Gamma/\dd q^2$ or $A_{\rm FB}$, or is independent of the
  Wilson coefficients. Data on the multitude of angular observables can hence be
  used to test our framework, that is, whether there are further operators
  beyond \refeq{smbasis}, the goodness of the OPE, and the form factors.
  
\item The $H_T^{(1,2,3)}$, by construction, do not depend on the form
  factors. Within our framework, these are the only observables with this
  feature, see Appendix \ref{sec:Ui}. 
  
  \item Moreover, $H_T^{(1)}$ does not depend on Wilson coefficients either.
  Its simple prediction \refeq{Hi:HQET} holds beyond the SM and provides a null test of the framework.
  
\item The set of observables Eqs.~(\ref{eq:dG:HQET})-(\ref{eq:Hi:HQET})  and (\ref{eq:ff-ratio-test}) with two short distance and three form factor
  coefficients is heavily overconstrained. Measurements can directly yield
  either products $\rho_i f_j f_k$ or ratios $\rho_2/\rho_1$ and $f_j/f_k$, but
  not the $f_i$ or the $\rho_{i}$ alone.
\end{itemize}

%
%--------+---------+---------+---------+---------+---------+---------+---------+
\section{Exploiting data \label{sec:SMdata}}

We give numerical SM predictions for $ \bar B \to \bar K^* l^+ l^-$ decay
observables in Section \ref{sec:smstatus}, with emphasis on the low recoil
region. In Section \ref{sec:fits} we confront the distributions with existing data and work out
constraints for the Wilson coefficients. Next, we combine low with large recoil regions and point out complementarities.

%
%--------+---------+---------+---------+---------+---------+---------+---------+
\subsection{SM predictions \label{sec:smstatus}}

\begin{table}
\begin{tabular}{|llr|llr|}
\hline \hline
$|V_{tb}V_{ts}^*|$        & $0.0409\pm0.0013$               &  &
$\alpha_s(M_Z)$          & $0.1176\pm0.0020$               & \cite{Amsler:2008zzb}
\\
$|V_{cb}|$                & $0.0417\pm0.0013$               &  &
$\alpha_e(m_b)$          & $1/133$                         &
\\
$|V_{ub}|/|V_{tb}V_{ts}^*|$ & $0.0884^{+0.064}_{-0.054}$      &  &
$\tau_{B^0}$              & $(1.530\pm0.009)~\pico\second$  & \cite{Amsler:2008zzb}
\\
$m_c(m_c)$ & $(1.27^{+0.07}_{-0.11})~\GeV$   & \cite{Amsler:2008zzb} & 
$f_{B^0}$                 & $(200\pm30)~\MeV$               & 
\\
$m_b(m_b)$ & $(4.2 \pm 0.17)~\GeV$           & \cite{Amsler:2008zzb} &
$f^{K^*}_{\parallel}$     & $(217\pm5)~\MeV$                & 
\\
$m_t^{\rm pole}$          & $(173.1\pm1.3)~\GeV$            & \cite{:2009ec} & 
$f^{K^*}_{\perp}(1 \GeV)$& $(185\pm10)~\MeV$               & 
\\
$M_W$                     & $(80.398\pm0.025)~\GeV$         & \cite{Amsler:2008zzb} &
$\lambda_{B,+}(1.5 \GeV)$& $(0.458 \pm 0.115)~\GeV$        & \cite{Beneke:2004dp}
\\
$M_Z$                     & $(91.1876\pm0.0021)~\GeV$       & \cite{Amsler:2008zzb} &  
$a_{1,K^*}^{\parallel,\perp}$  & $(0.1 \pm 0.07)~\MeV$      & \cite{Ball:2004rg}
\\
$\mathcal{B}(\bar{B}\to X_cl\bar{\nu}_l)$ & $(10.5\pm0.4)~\%$& \cite{Amsler:2008zzb} & 
$a_{2,K^*}^{\parallel,\perp}$  & $(0.1 \pm 0.1)~\MeV$       & \cite{Ball:2004rg}
\\
\hline \hline
\end{tabular}
\caption{The numerical input used in our analysis.
  We neglect the mass of the strange quark.
   $\tau_{B^0}$ denotes
the lifetime of the neutral $B$ meson.
  \label{tab:numericConstants}
}
\end{table}

The low recoil predictions are obtained using the formulae given in Section
\ref{sec:loreco}.  The framework applies to the region where the $\bar K^*$ is
soft in the heavy mesons rest frame, {\it i.e.}, has energy $E_{K^*}=m_{K^*} +
\Lambda$. In terms of dilepton masses, this corresponds to large values, $q^2
\gtrsim (m_B - m_{K^*})^2 - 2 m_B \Lambda$ up to the kinematical endpoint.
We use, unless otherwise stated,
\begin{align} \label{eq:loreco-range}
q^2_{\rm min}=14 \GeV^2 < q^2 \leq 19.2 \GeV^2 =q^2_{\rm max} ,
\end{align}
obtained numerically for $\Lambda= 500 \, \rm
MeV$, with the lower boundary starting just above the $\psi^\prime$ resonance.

To make quantitative predictions in the low recoil region the $B \to K^*$ form factors are requisite input. Unfortunately, the current knowledge on the form factors at low recoil is very limited and
our results can as far as form factor uncertainties are concerned provide guidance of the achievable precision only.

For our numerics we use  the light cone sum rule (LCSR)
results  of Ref.~\cite{Ball:2004rg} extrapolated from their domain of validity at large recoil to the low recoil one with physical pole or dipole shapes. These extrapolations 
are supported by fits based on series expansion in the case of $B\to K$
and $B \to\rho$ transitions \cite{Bharucha:2010im}. 
Note that there is lattice and
experimental information available on $B \to \rho$ form factors at low recoil
\cite{Abada:2002ie,Flynn:2008zr}, however, to use this for $B \to K^*$ would
require  knowledge of the size of $SU(3)$ flavor breaking.
More details on the form factors and a
comparison with existing lattice results for
$T_{1,2}$ \cite{Becirevic:2006nm, Liu:2009dj} are given in Appendix \ref{sec:ChoiceFF}.
We use the parameters given in \reftab{numericConstants}. 

From $q^2$-integration in the low recoil region \refeq{loreco-range}
we obtain the integrated SM branching ratio $\mathrm{d}\mathcal{B}/\mathrm{d}q^2 = \tau_{B}
\mathrm{d}\Gamma/\mathrm{d}q^2$ as
\begin{align} \label{eq:loreco-SMBR}
  10^7\cdot \int_{q^2_{\rm min}}^{q^2_{\rm max}} 
    \dd q^2 \frac{\dd \cal {B}}{\dd q^2} & = 
   2.96\, {^{+0.90}_{-0.77}}\Big|_{\rm FF}\, {^{+0.18}_{-0.17}}\Big|_{\rm SL}\, 
          {\pm 0.10}\Big|_{\rm IWR}\, {\pm 0.16}\Big|_{\rm CKM} \, {^{+0.08}_{-0.06}}\Big|_{\rm SD}.
\end{align}
For the remaining $q^2$-distributions $X$ of 
Eqs.~(\ref{eq:Afb})-(\ref{eq:def:HT3}) we define ``naively integrated'' observables as
\begin{align} 
   \overline{X} \equiv \frac{1}{q^2_{\rm max} - q^2_{\rm min}} 
   \int_{q^2_{\rm min}}^{q^2_{\rm max}} \dd q^2 X(q^2)\,.
\end{align}
For these we obtain
\begin{align}
  \overline{A_{\rm FB}^{\phantom{L}}} & = -0.39 \, 
                  {^{+0.06}_{-0.07}}\Big|_{\rm FF} \, 
                  {\pm 0.02}\Big|_{\rm SL} \, 
                  {\pm 0.01}\Big|_{\rm IWR}\, {\pm 0.001}\Big|_{\rm SD},
\\
  \overline{F_{\rm L}^{\phantom{L}}} & = 0.35 \, 
                  {^{+0.03}_{-0.04}}\Big|_{\rm FF} \, 
                  {\pm 0.03}\Big|_{\rm SL} \, 
                  {\pm 0.01}\Big|_{\rm IWR}\,,
\\
  \overline{A_T^{(2)}} & = -0.54 \, 
                  {^{+0.15}_{-0.13}}\Big|_{\rm FF} \,
                  {^{+0.04}_{-0.03}}\Big|_{\rm SL} \, 
                  {^{+0.03}_{-0.02}}\Big|_{\rm IWR}\,,
\\
  \overline{A_T^{(3)}} & = +2.25 \, 
                  {^{+0.52}_{-0.45}}\Big|_{\rm FF} \,
                  {\pm 0.11}\Big|_{\rm SL} \, 
                  {\pm 0.08}\Big|_{\rm IWR}\,,
\\
  \overline{A_T^{(4)}} & = +0.53 \,
                  {^{+0.12}_{-0.11}}\Big|_{\rm FF} \,
                  {\pm 0.03}\Big|_{\rm SL} \,
                  {\pm 0.02}\Big|_{\rm IWR}\, {\pm 0.002}\Big|_{\rm SD},
\\
  \overline{H_T^{(1)}} & = +1.000 \,
                  {^{+0}_{-0.00002}}\Big|_{\rm SL} \,
                  {^{+0}_{-0.0007}}\Big|_{\rm IWR}\,,
\\
  \overline{H_T^{(2,3)}} & = -0.985 \,
                  {\pm 0.001}\Big|_{\rm SL} \, 
                  {^{+0.007}_{-0.005}}\Big|_{\rm IWR}\,
                  {^{+0.004}_{-0.003}}\Big|_{\rm SD}.
\end{align}

In addition, we consider the integrated observables  $\langle X\rangle$ defined
by replacing $J_i$ with its integral $\langle J_i \rangle$
in each of the observables $X=X(J_i)$ in Eqs.~(\ref{eq:dGam:def})-(\ref{eq:def:HT3}),
\begin{align}
 \langle X\rangle \equiv X( \langle J_i  \rangle ) , ~~~~~~\langle J_i\rangle \equiv 
 \int^{q^2_{\rm max}}_{q^2_{\rm min}} \dd q^2 J_i(q^2) \label{eq:integratedObservables}.
\end{align}
This definition agrees with the way ${\cal B}, A_{\rm
FB}$ and $F_{\rm L}$ are obtained experimentally \cite{:2008ju,:2009zv,CDF2010}, {\it i.e.}, by integrating numerator and
denominator before taking the ratio. Using the same integration boundaries as above, \refeq{loreco-range}, we obtain
\begin{align}
  \langle A_{\rm FB} \rangle & = -0.41\, \label{eq:integratedafb}
                {\pm 0.07}\Big|_{\rm FF}\,
                {\pm 0.02}\Big|_{\rm SL}\,
                {\pm 0.01}\Big|_{\rm IWR}\,
                {^{+0.001}_{-0.002}}\Big|_{\rm SD},
\\
  \langle F_{\rm L} \rangle & = 0.35\,
                {^{+0.04}_{-0.05}}\Big|_{\rm FF}\,
                {\pm 0.03}\Big|_{\rm SL}\,
                {\pm 0.02}\Big|_{\rm IWR},
\\
  \langle A_T^{(2)} \rangle & = -0.48
                {^{+0.18}_{-0.15}}\Big|_{\rm FF}\,
                {^{+0.04}_{-0.04}}\Big|_{\rm SL}\,
                {\pm 0.03}\Big|_{\rm IWR}\,
                {\pm 0.001}\Big|_{\rm SD},
\\
  \langle A_T^{(3)} \rangle & = 1.71 \,
                {^{+0.39}_{-0.33}}\Big|_{\rm FF}\,
                {\pm 0.08}\Big|_{\rm SL}\,
                {\pm 0.06}\Big|_{\rm IWR}\,
                {\pm 0.001}\Big|_{\rm SD},
\\
  \langle A_T^{(4)} \rangle & = 0.58 \,
                {^{+0.13}_{-0.11}}\Big|_{\rm FF}\,
                {^{+0.09}_{-0.09}}\Big|_{\rm SL}\,
                {\pm 0.07}\Big|_{\rm IWR}\,
                {\pm 0.002}\Big|_{\rm SD},
\\
  \langle H_T^{(1)} \rangle &  = +0.997\,
                {\pm 0.002}\Big|_{\rm FF}\,
                {^{+0}_{-0.001}}\Big|_{\rm IWR},
\\
  \langle H_T^{(2)} \rangle & =  -0.972\,
                {^{+0.004}_{-0.003}}\Big|_{\rm FF}\,
                {\pm 0.001}\Big|_{\rm SL}\,
                {^{+0.008}_{-0.005}}\Big|_{\rm IWR}\,
                {^{+0.003}_{-0.004}}\Big|_{\rm SD},
\\
  \langle H_T^{(3)} \rangle & =  -0.958\, \label{eq:integratedat7}
                {\pm 0.001}\Big|_{\rm SL}\,
                {^{+0.008}_{-0.006}}\Big|_{\rm IWR}\,
                {^{+0.003}_{-0.004}}\Big|_{\rm SD},
\end{align}
with the branching ratio as before, \refeq{loreco-SMBR}.
Uncertainties not explicitly given are below ${\cal{O}}(10^{-4})$.

In both definitions of integrated observables the uncertainties are estimated the same way: 
The dominant uncertainty of the
form factors $V$, $A_1$ and $A_2$ has been assumed $\pm 15$~\%
(FF). Furthermore, we include a real scaling factor for each of the transversity
amplitudes $A_{\perp,\parallel,0}^{L,R}$ in order to estimate uncertainties due
to the subleading corrections of order $\alpha_s \Lambda/m_b$ by varying them with
$\pm 5$~\% (SL). The subleading corrections to the improved Isgur-Wise form
factor relations \refeq{ffrelations}, of order $\Lambda/m_b$, and the neglected
kinematical factors of $m_{K^*}/m_B$ in the term $\sim \kappa\, \wilson[eff]{7}$
are accounted for by three real scale factors for $A_{\perp,\parallel,0}$ with $\pm
20$~\% (IWR). Note however, that the latter are additionally suppressed in the SM by $2\,
\wilson[eff]{7}/ \wilson[eff]{9}$. The uncertainties due to the CKM
parameters $V_{tb} V_{ts}^*$ correspond to their $1 \sigma$ ranges (CKM), which cancel in the
normalized quantities and thus appear  in the branching ratio only.
The uncertainties due to the $\mu$-dependence and the $t$- and $b$-quark masses
(at 1 $\sigma$)
concern the short distance couplings $\rho_{1,2}$ only, and are subsumed under the label (SD). 
The variation with the scale $\mu \in
[\mu_b/2, 2\mu_b] $ (with central value $\mu_b = 4.2\, \GeV$) is small, as expected. 
\begin{figure}
\centering
{\includegraphics[angle=-90,width=.45\textwidth]{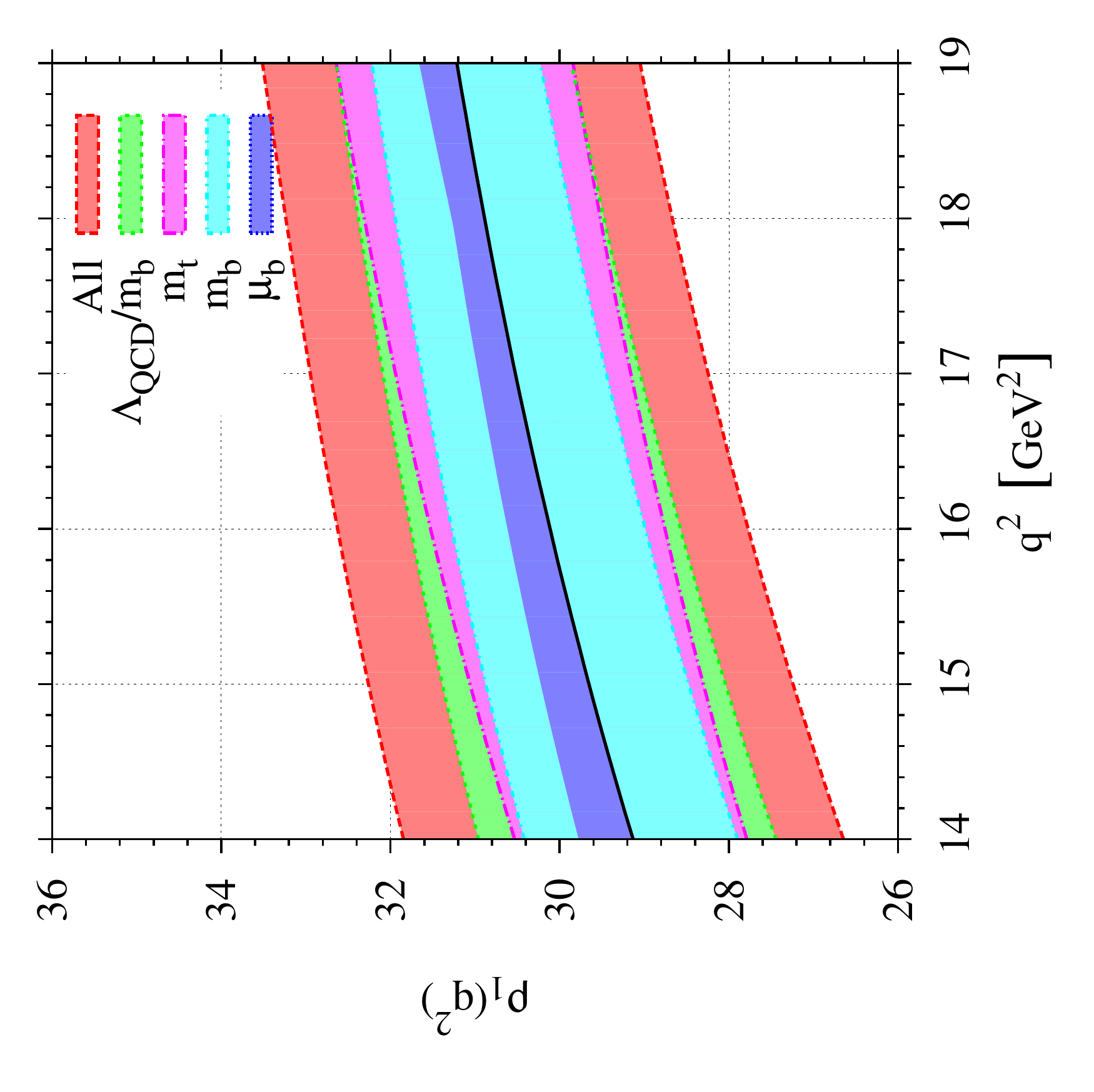}}
{\includegraphics[angle=-90,width=.45\textwidth]{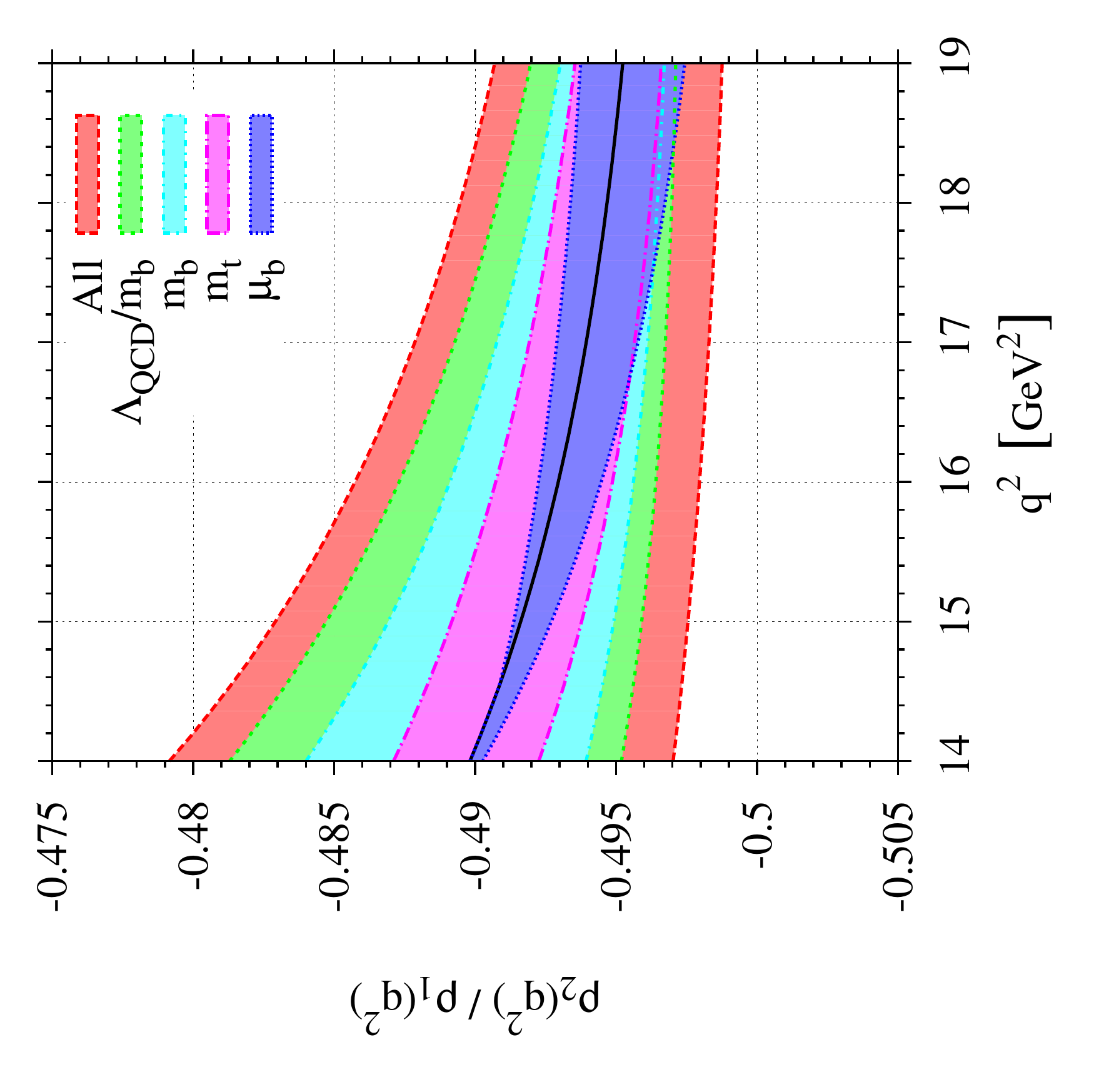}}
\caption{The short distance coupling $\rho_1$ and the
  ratio $\rho_2/\rho_1$ in the SM.}  
\label{fig:rho-in-SM}
\end{figure}

\begin{figure}
\centering
\subfigure[]{\includegraphics[angle=-90,width=.7\textwidth]{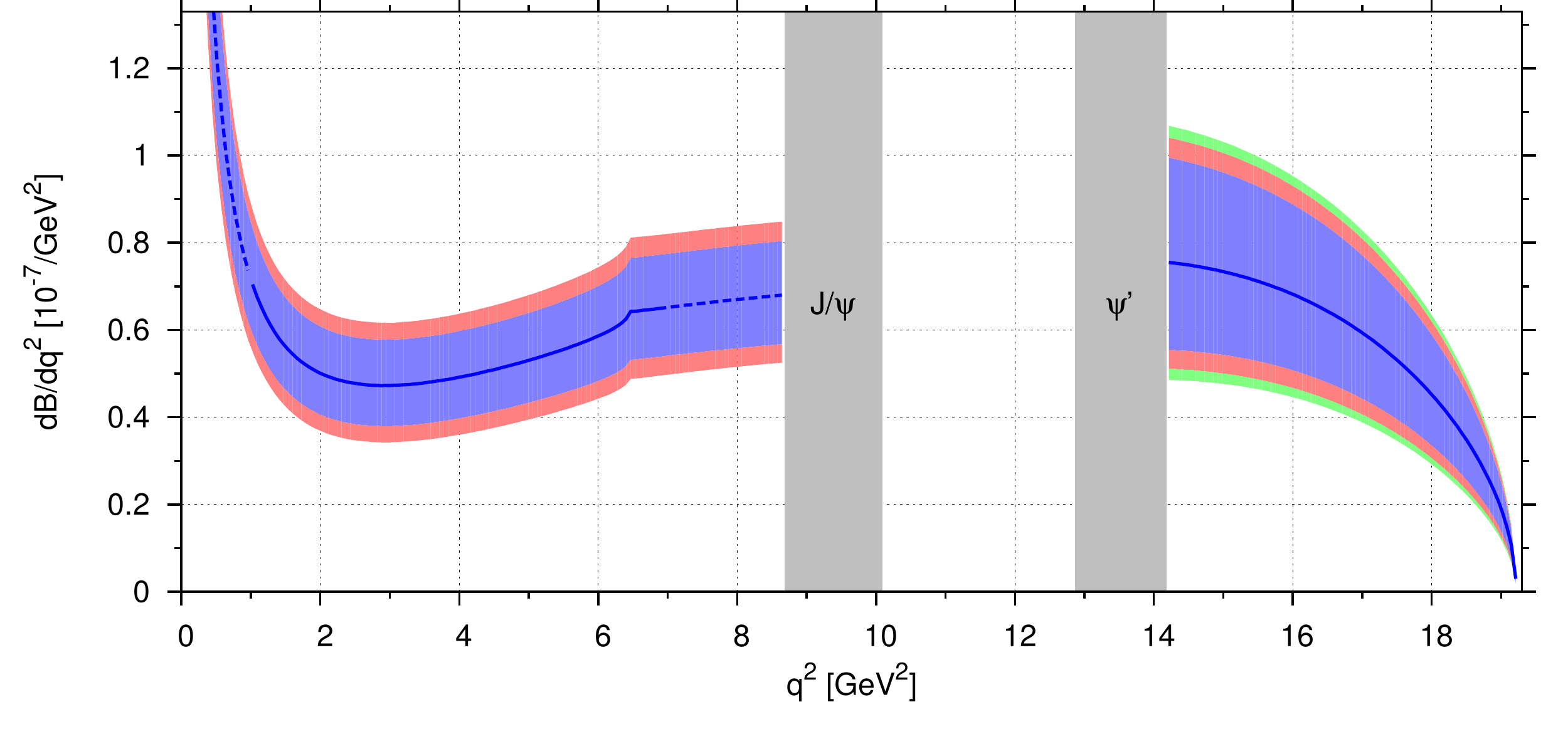}}\\
\subfigure[]{\includegraphics[angle=-90,width=.7\textwidth]{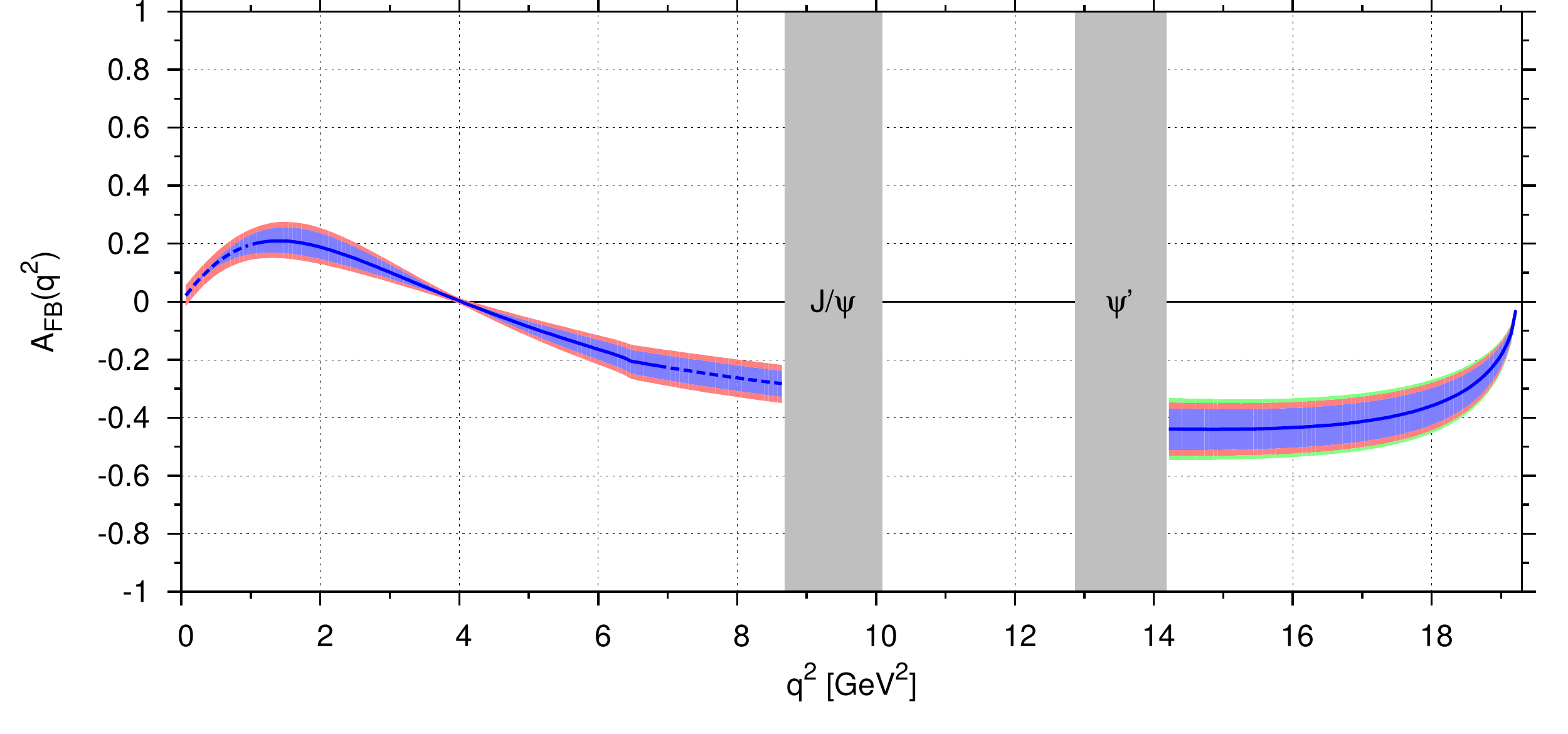}}\\
\subfigure[]{\includegraphics[angle=-90,width=.7\textwidth]{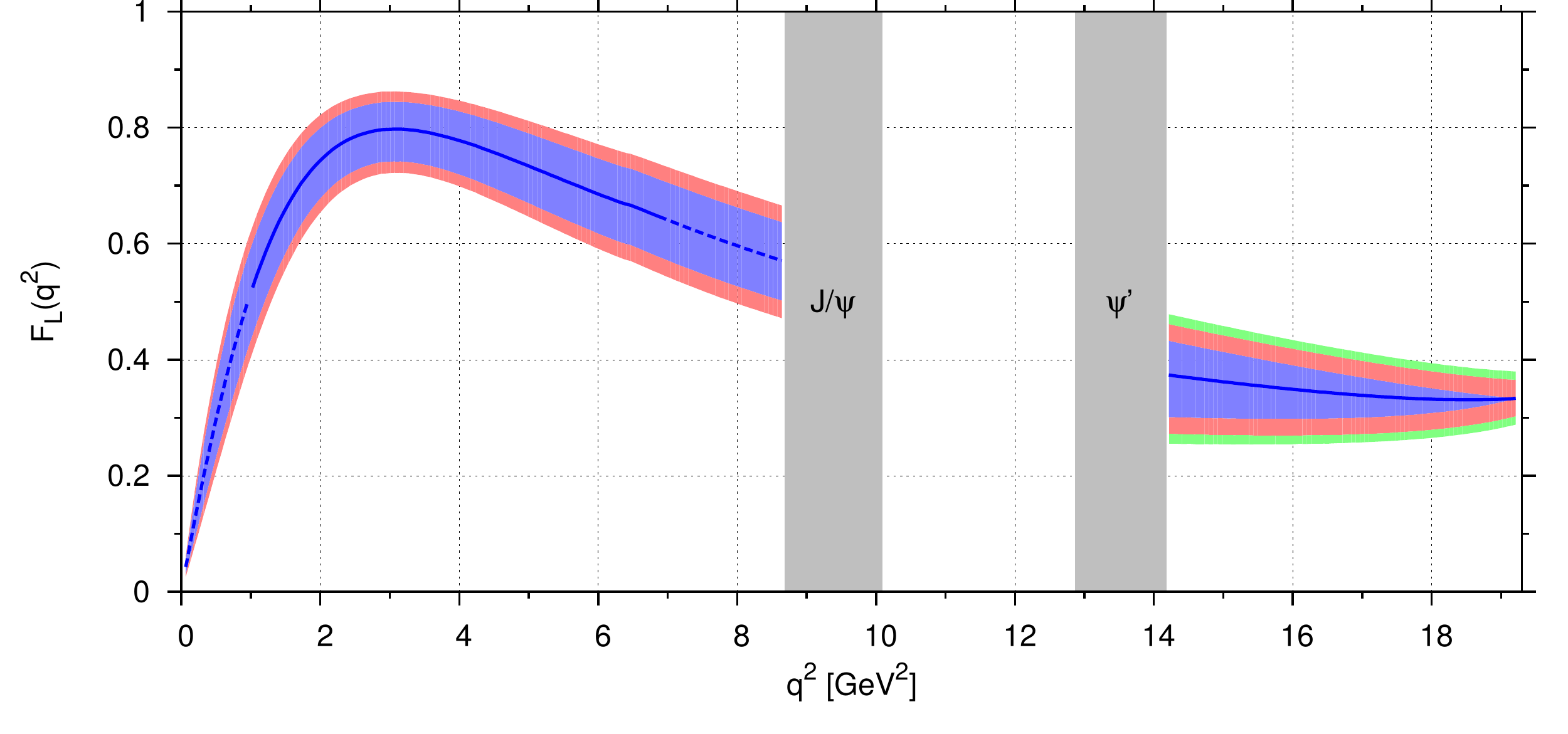}}
\caption{The differential branching ratio $\mathrm{d}\mathcal{B}/\mathrm{d}q^2$
  in units of $10^{-7}/\rm GeV^2$ (a), the forward-backward asymmetry $A_{\rm
  FB}$ (b) and the longitudinal polarization $F_{\rm L}$ (c) in the large
  recoil $q^2 < m_{J/\psi}^2$ and the low recoil $q^2 \gtrsim m_{\psi^\prime}^2
  \sim {\cal{O}}(m_b^2)$ regions in the SM.  At low recoil, 
  the uncertainties shown are due to the $\Lambda/Q$ expansion of the improved
  Isgur-Wise relations (green bands), subleading terms of order
  $\alpha_s\Lambda/Q$ (red bands) and the form factors (blue bands).  At large
  recoil, the bands denote the uncertainties from $\Lambda/m_b,\,
  \Lambda/E_{K^*}$ corrections (red bands) and the form factors (blue bands).
  The vertical shaded (grey) bands mark
  the experimental veto regions \cite{:2009zv,CDF2010} to remove contributions
  from $\bar B \to J/\psi (\to \mu^+ \mu^-) \bar K^*$ (left band) and $\bar B
  \to \psi^\prime (\to \mu^+ \mu^-) \bar K^*$ (right band).  }
\label{fig:2observables-in-SM}
\end{figure}

\begin{figure}
\centering
\subfigure[]{\includegraphics[angle=-90,width=.7\textwidth]{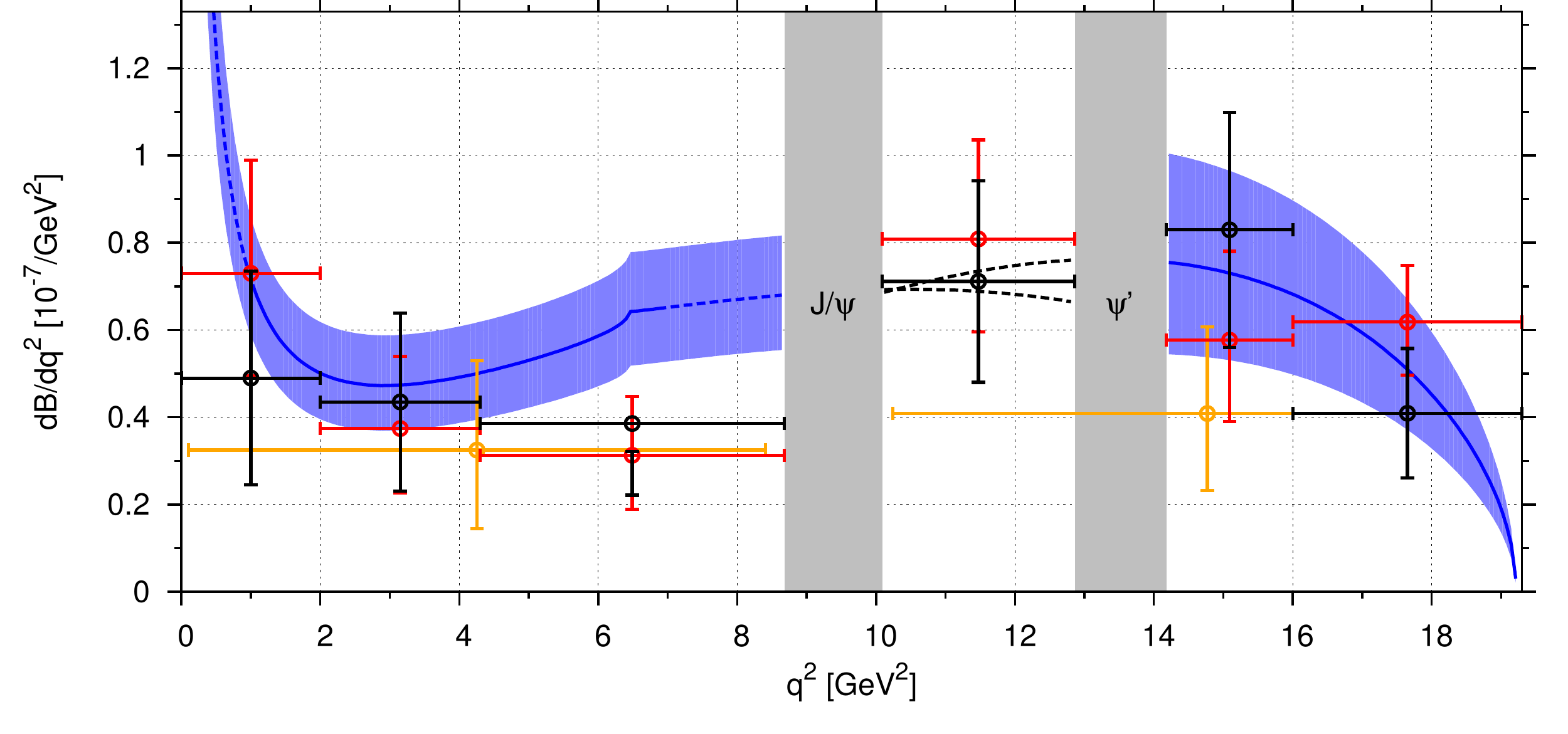}}\\
\subfigure[]{\includegraphics[angle=-90,width=.7\textwidth]{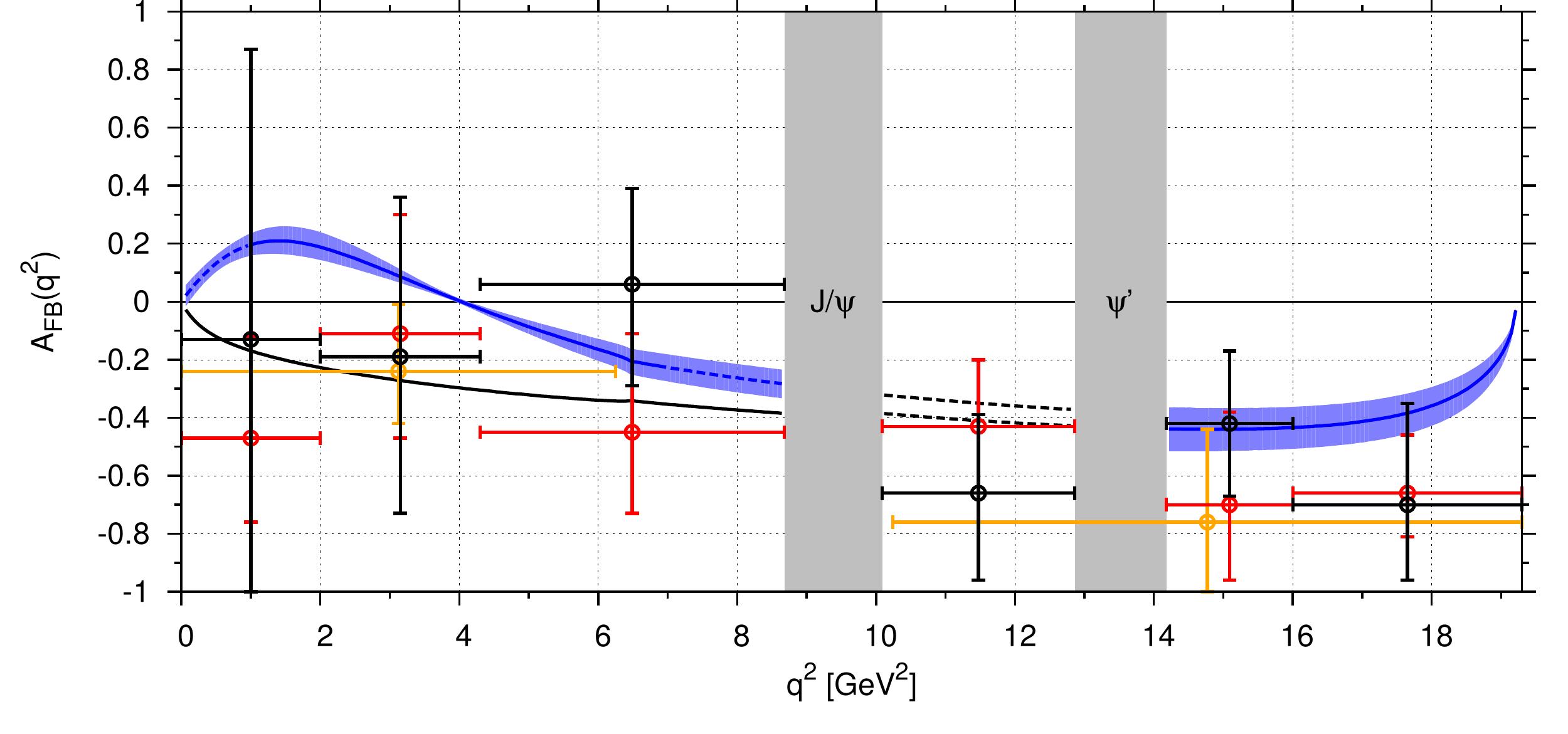}}\\
\subfigure[]{\includegraphics[angle=-90,width=.7\textwidth]{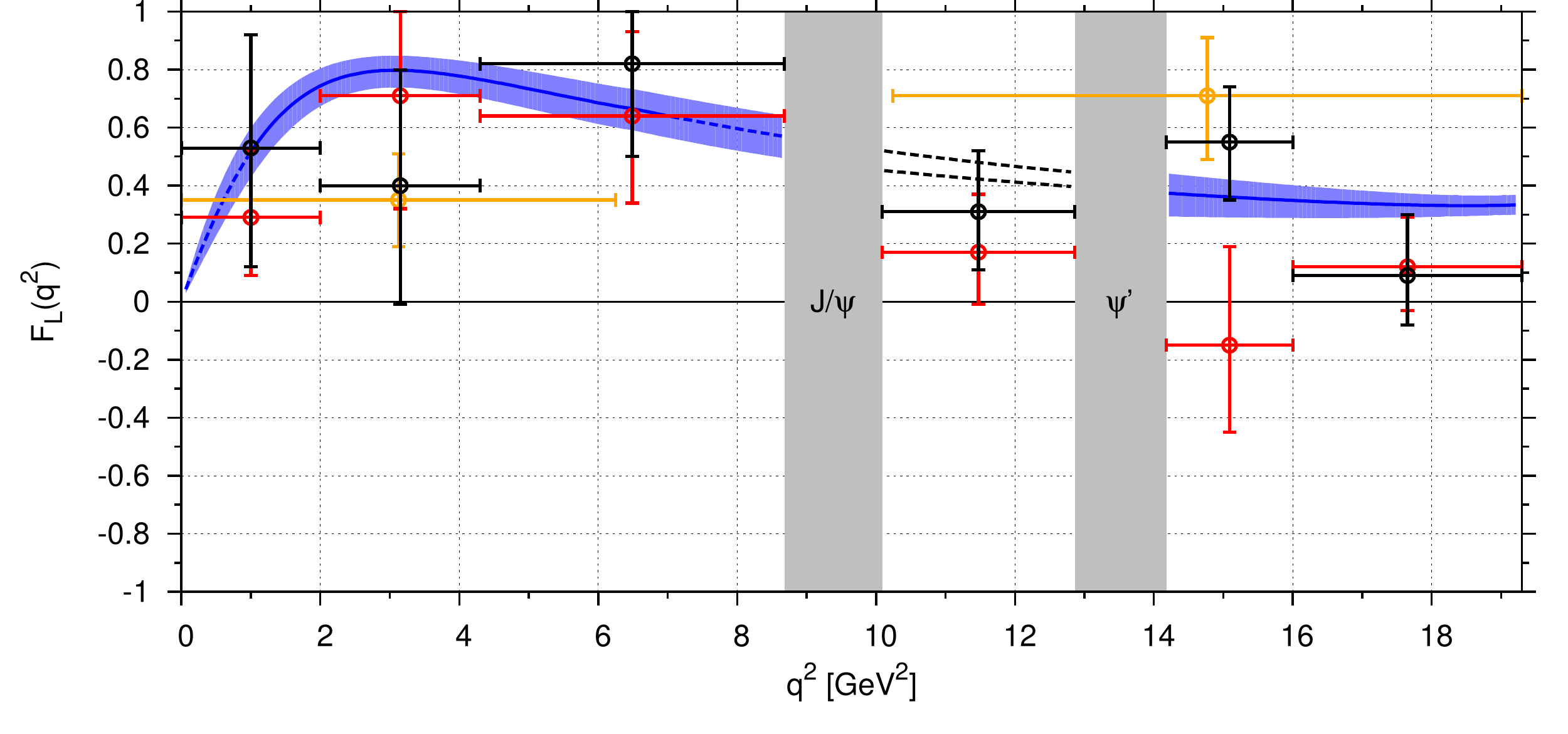}}
\caption{The $\bar B \to \bar K^* l^+ l^-$ distributions $\mathrm{d}\mathcal{B}/\mathrm{d}q^2$ (a), $A_{\rm
  FB}$ (b) and $F_{\rm L}$ (c) in the SM including
  the theoretical uncertainties added in quadrature (shaded blue bands) versus the
  existing data from Belle \cite{:2009zv} (red), BaBar
  \cite{Aubert:2006vb,:2008ju} (gold) and CDF \cite{CDF2010} (black).  The
  experimental data for $A_{\rm FB}$ have their sign flipped to match the
  conventions used in this work.  The isolated solid (black) line in the $A_{\rm FB}$ plot
  illustrates the case with ${\cal{C}}_7 =-{\cal{C}}_7^{\rm SM}$. The vertical shaded (grey) bands are defined as
  in \reffig{2observables-in-SM}. The isolated dashed (black) lines between the
  $ \bar c c$-bands are theory extrapolations from the low and large recoil region.}
\label{fig:observables-in-SM}
\end{figure}

\begin{figure}
\centering
\subfigure[]{\includegraphics[angle=-90,width=.7\textwidth]{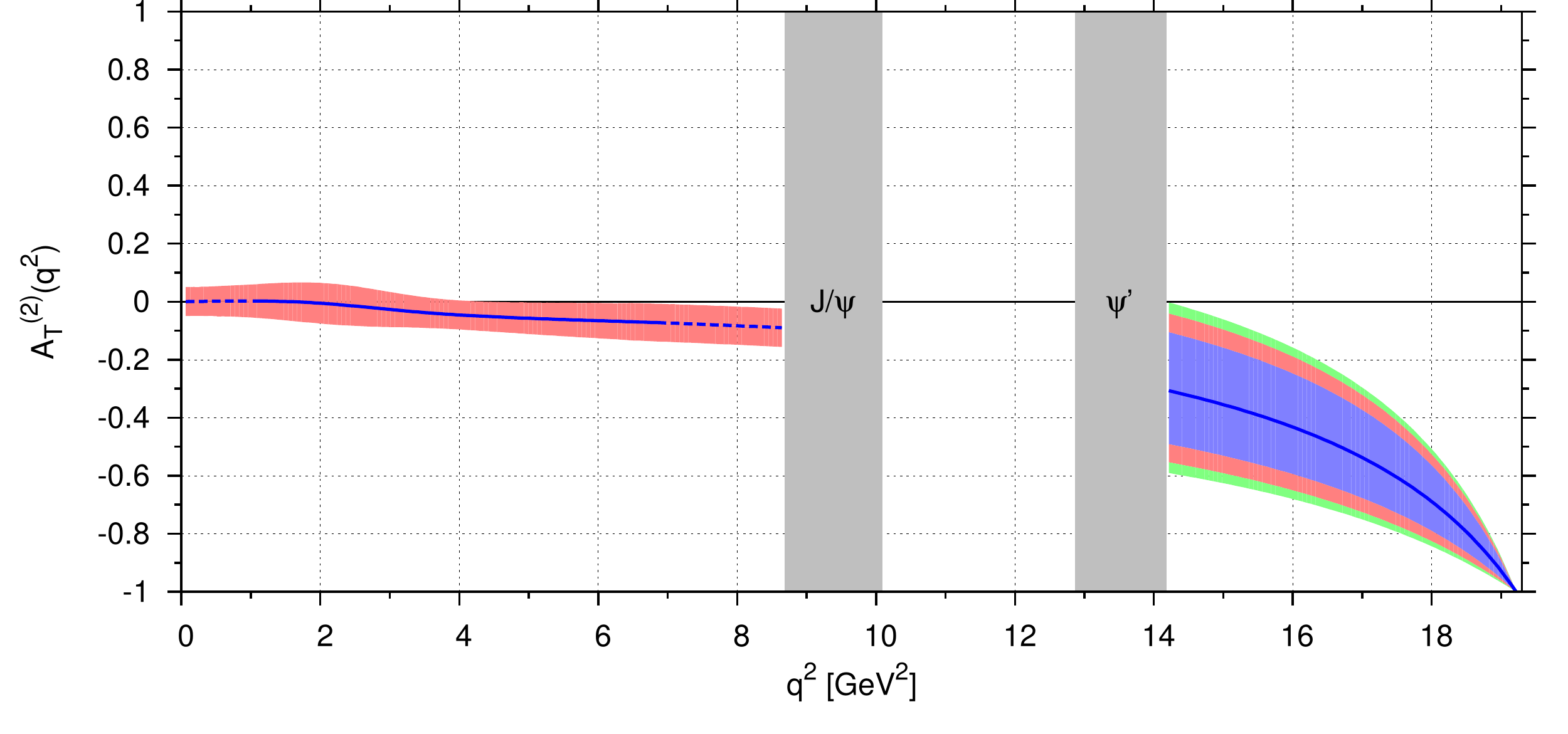}}
\\ 
\subfigure[]{\includegraphics[angle=-90,width=.7\textwidth]{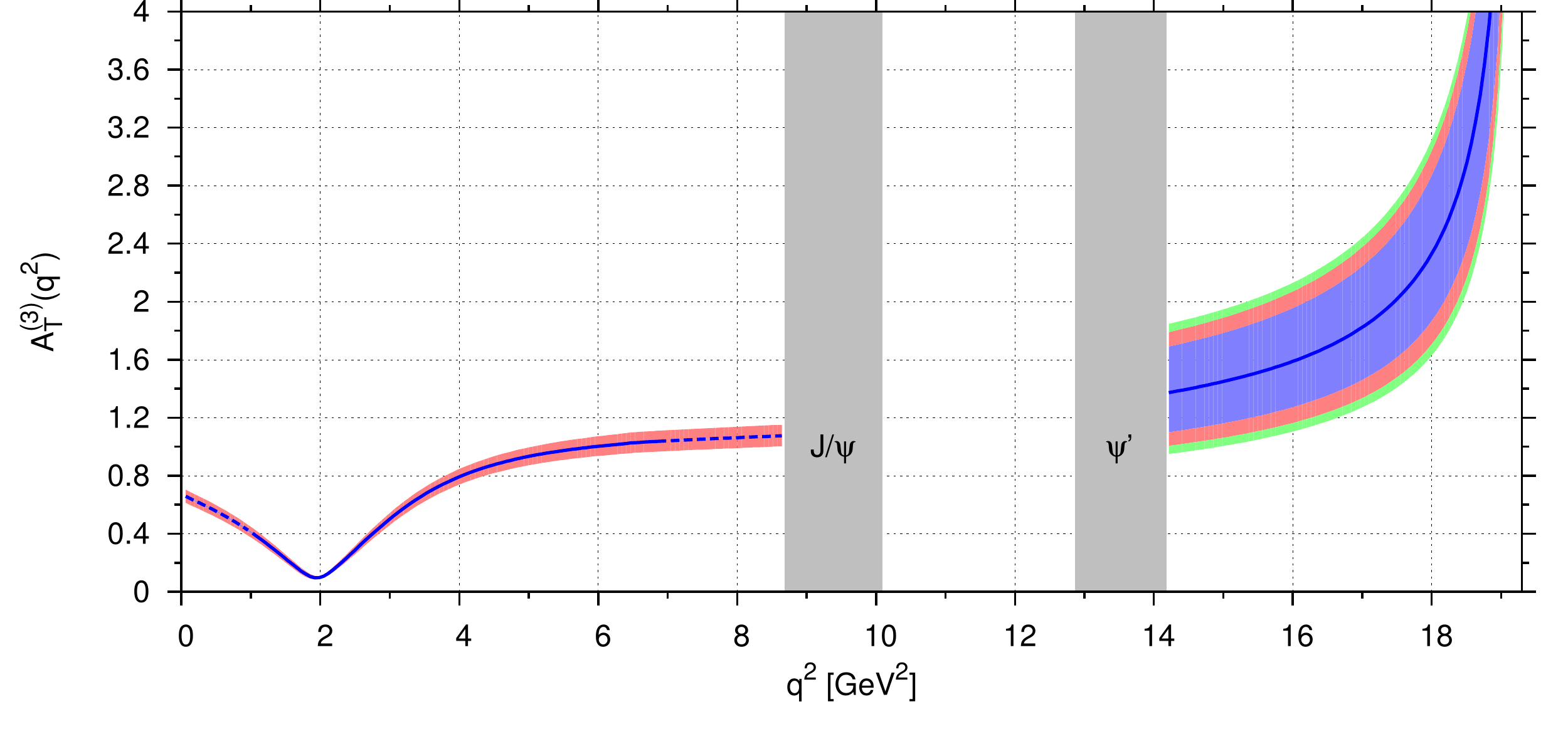}}
\\   
\subfigure[]{\includegraphics[angle=-90,width=.7\textwidth]{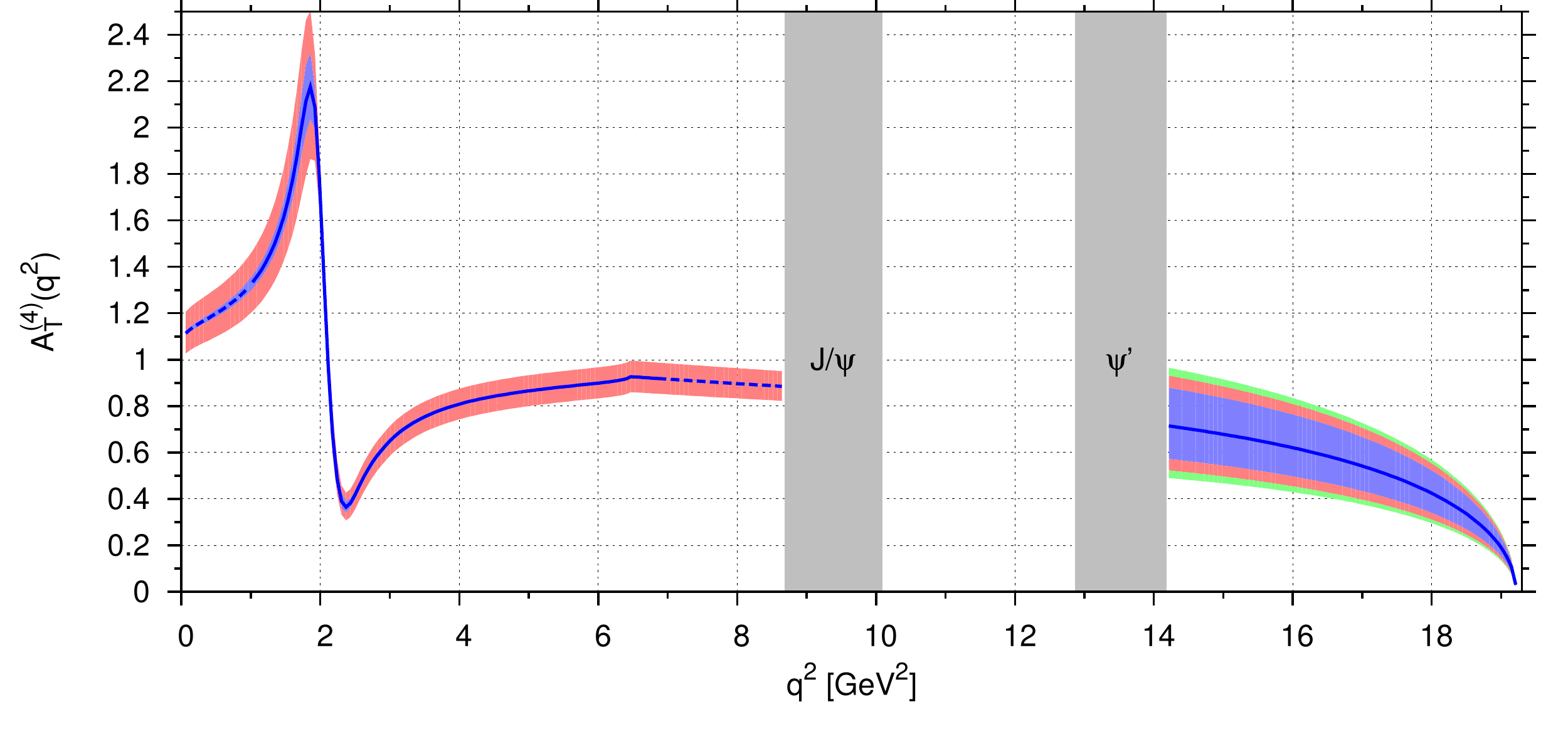}}
\caption{The transverse asymmetries $A_T^{(2)}$ (a), $A_T^{(3)}$ (b) and
  $A_T^{(4)}$ (c) in the SM. The explanation of the bands is the same as in
  \reffig{2observables-in-SM}.}
\label{fig:at234}
\end{figure}

In \reffig{rho-in-SM} we show $\rho_1$ and the ratio $\rho_2/\rho_1$ with error bands 
from different sources. 
The $t$-pole mass
and $b$-MS mass dependence (at $3 \sigma$) are comparable in size and amount to
about $5$~\% each. Finally, a variation of $\pm 20$~\% due to the subleading
$\Lambda/m_b$ and $m_{K^*}/m_B$ corrections denoted above as (IWR) results in
about $6$~\% uncertainty. The overall uncertainty of $\rho_{1}$ and $\rho_2$ is about $9$~\%
and $10$~\%, respectively, when adding all uncertainties in quadrature.
However, the uncertainties cancel to a large extent in the ratio $\rho_2/\rho_1$,
providing a strong test of the SM when measuring the observables $H_T^{(2,3)}$
[\refeqs{def:HT2}{def:HT3}] with an uncertainty of about $2$~\% (at $3\sigma$).

The uncertainties of each group (FF), (SL), (IWR), (CKM) and (SD) are obtained by varying
each parameter separately and adding them subsequently in quadrature. 
Our SM values for $\langle A_{\rm FB} \rangle$ and $\langle F_{\rm L}\rangle$
are in agreement with \cite{Bauer:2009cf}.

For the SM predictions at large recoil \cite{Beneke:2001at,Beneke:2004dp} we
follow closely \cite{Bobeth:2008ij}, with the updates of the numerical input
given in \reftab{numericConstants}. 
In this kinematical region, spectator effects arise and for concreteness, we give
predictions for neutral $\bar B$ decays.

We estimate the uncertainties due to the two large energy form 
factors $\xi_{\perp,\parallel}$ by
varying them separately -- for an improved treatment of this source of
uncertainty using directly the LCSRs  the reader is refered to \cite{Altmannshofer:2008dz}.
Furthermore, we estimate uncertainties due to subleading QCDF corrections of
order $\Lambda/m_b$ by varying a real scale factor for each of the transversity
amplitudes $A_{\perp,\parallel,0}^{L,R}$ within $\pm 10$~\% separately and
adding the resulting uncertainties subsequently in quadrature. The latter 
constitute the numerically leading uncertainties in the observables $A_T^{(2,3,4)}$ where form factor uncertainties cancel at leading order in QCDF \cite{Egede:2008uy}.

The differential branching ratio $\mathrm{d}\mathcal{B}/\mathrm{d}q^2$, the forward-backward asymmetry $A_{\rm FB}$ and
the longitudinal polarization $F_{\rm L}$ in the SM in both the low and large
recoil regions are shown in \reffig{2observables-in-SM}.  
The vertical grey bands are the regions vetoed by
the experiments to remove backgrounds from intermediate charmonia, $J/\psi$ and
$\psi^\prime$ decaying to muon pairs for $8.68 \, {\rm GeV}^2< q^2 < 10.09\, {\rm GeV}^2$ and
$12.86 \, {\rm GeV}^2 < q^2 < 14.18 \, {\rm GeV}^2$ \cite{:2009zv,CDF2010}.  
Within QCDF, the
region of validity is approximately within $(1 - 7) \, \mbox{GeV}^2$.
We mark the large recoil range (below the $J/\psi$) outside
this range by dashed lines.

In \reffig{observables-in-SM} we show the SM predictions for $\mathcal{B}, A_{\rm FB}$
and $F_{\rm L}$ next to the available data. Note that the physical region
of $F_{\rm L}$ is between 0 and 1. The data are consistent with the SM, although they
allow for large deviations from the SM as well given the sizeable uncertainties.
In particular, the data for $\mathcal{B}$ at low $q^2$ and $A_{\rm FB}$ at high $q^2$
show a trend to be slighly below the SM. The shape of $A_{\rm FB}$ at low $q^2$ is currently
not settled and allows for either sign of the dipole coefficient ${\cal{C}}_7$ while
having the others kept at their SM values.
In the future the LHC$b$ collaboration expects to surpass the precision of the existing $B$-factory
$A_{\rm FB}$ measurements after an integrated luminosity of $0.3\, {\rm
fb}^{-1}$ \cite{:2009ny}, and may shed light on this matter.

In \reffig{at234} we show $A_T^{(2,3,4)}$ in the SM. The behaviour in the low
and high $q^2$ region is very different from each other. In particular, $A_T^{(2)}$ is
strongly suppressed, in fact, vanishes up to $1/E_{K^*}$ corrections by helicity
conservation \cite{Burdman:2000ku} for low dilepton masses, but is order one for
large ones. The size of  $A_T^{(2)}$ at low $q^2$ 
can be used as an indicator for the correctness of our assumptions: in the presence of chirality-flipped operators beyond those in \refeq{smbasis}, the aforementioned suppression of 
$A_T^{(2)}$ would be lifted.
Note that $A_T^{(3)}$ is proportional to $1/\sqrt{\hat{\lambda}}$ and 
diverges at the endpoint $\hat{\lambda} \to 0$. On the other hand, $A_T^{(4)}
\propto \sqrt{\hat{\lambda}}$ is finite in this limit and vanishes at maximum $q^2$.

The $q^2$-behaviour of both the new, transverse observables $H_T^{(2,3)}$ can be obtained from  \reffig{rho-in-SM}, where $\rho_2/\rho_1$ is shown in the SM.

%
%--------+---------+---------+---------+---------+---------+---------+---------+
\subsection{Constraining new physics  \label{sec:fits}}

\begin{figure}[b]
\begin{center}
  \subfigure[]{\includegraphics[angle=-90,width=.49\textwidth]{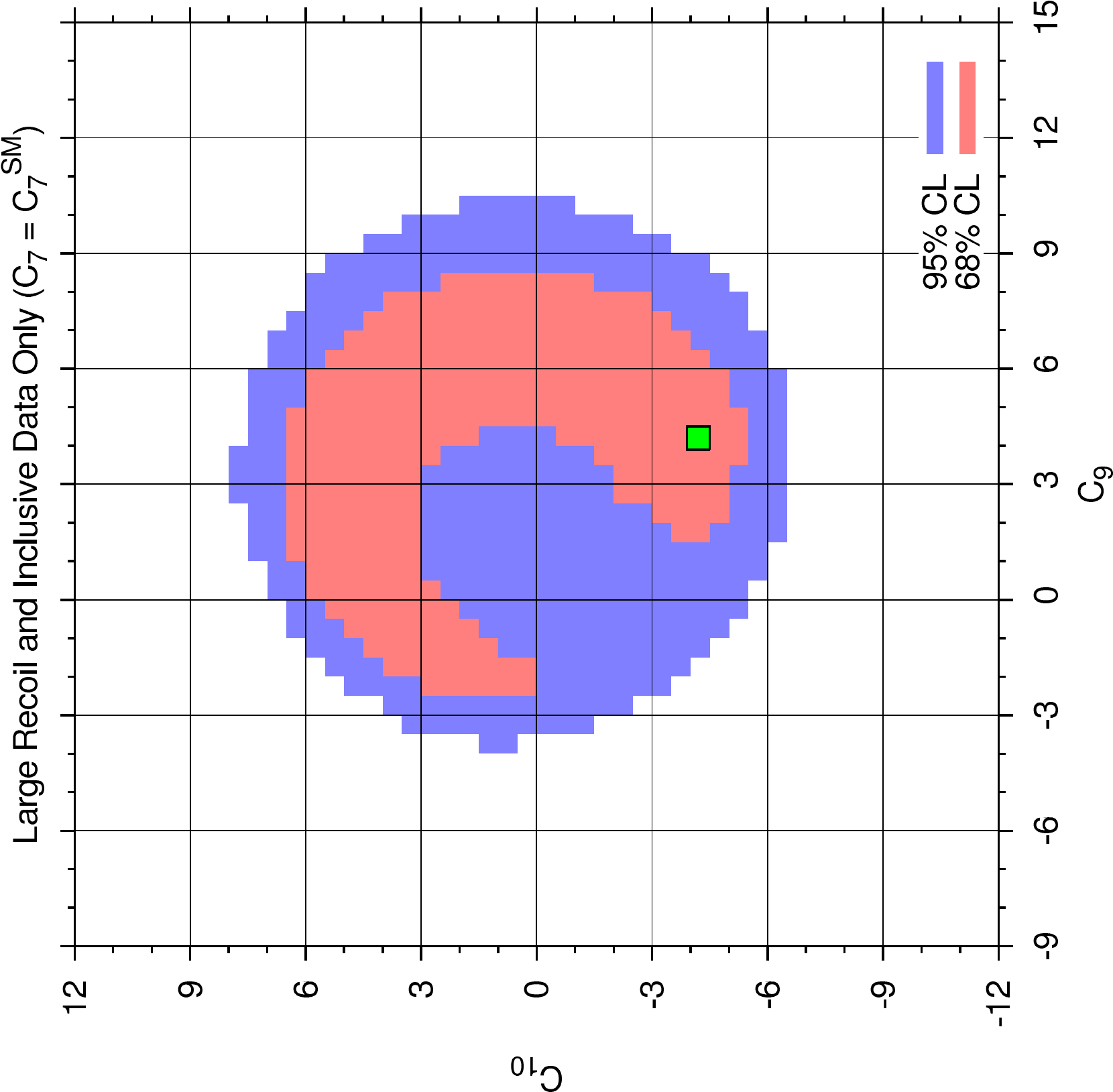}}
  \subfigure[]{\includegraphics[angle=-90,width=.49\textwidth]{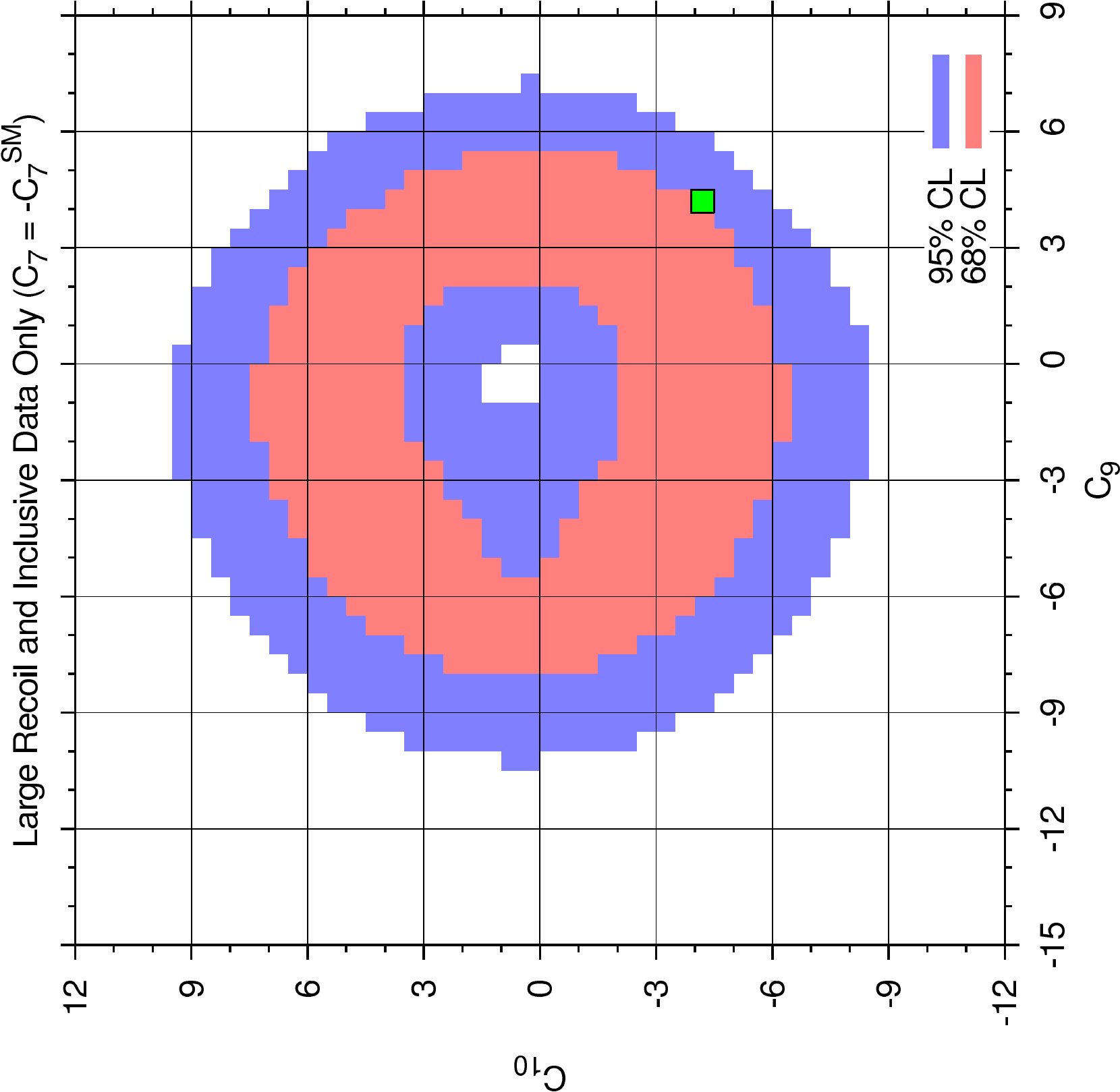}}
\end{center}
\caption{The constraints on $\wilson{9}$ and $\wilson{10}$
  from $\bar{B}\to \bar{K}^*l^+l^-$ at large recoil and
  $\bar{B}\to X_s l^+l^-$  for $\wilson{7} = \wilson[SM]{7}$ (a)
  and  $\wilson{7} = -\wilson[SM]{7}$ (b) using  Belle \cite{:2009zv,Iwasaki:2005sy}, BaBar \cite{Aubert:2004it} and CDF \cite{CDF2010} data at $68\%$~CL (red areas)
  and $95\%$~CL (red and blue areas).
  The (green) square marks the SM value of ($\wilson{9},\wilson{10}$).}
\label{fig:constraintsLowQ2+Inclusive}
\end{figure}

\begin{figure}[b]
\begin{center}
  \subfigure[]{\includegraphics[angle=-90,width=.49\textwidth]{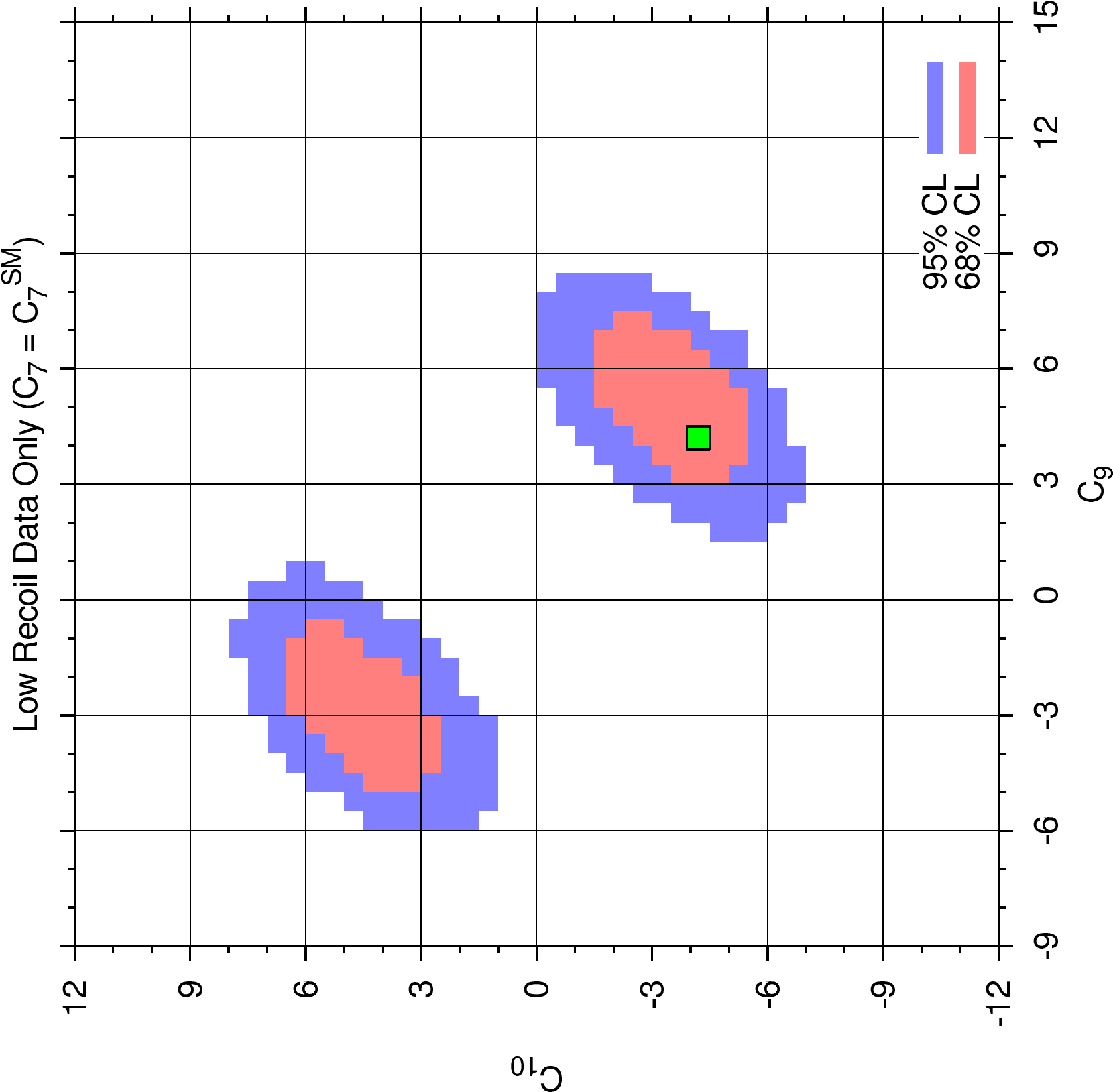}}
  \subfigure[]{\includegraphics[angle=-90,width=.49\textwidth]{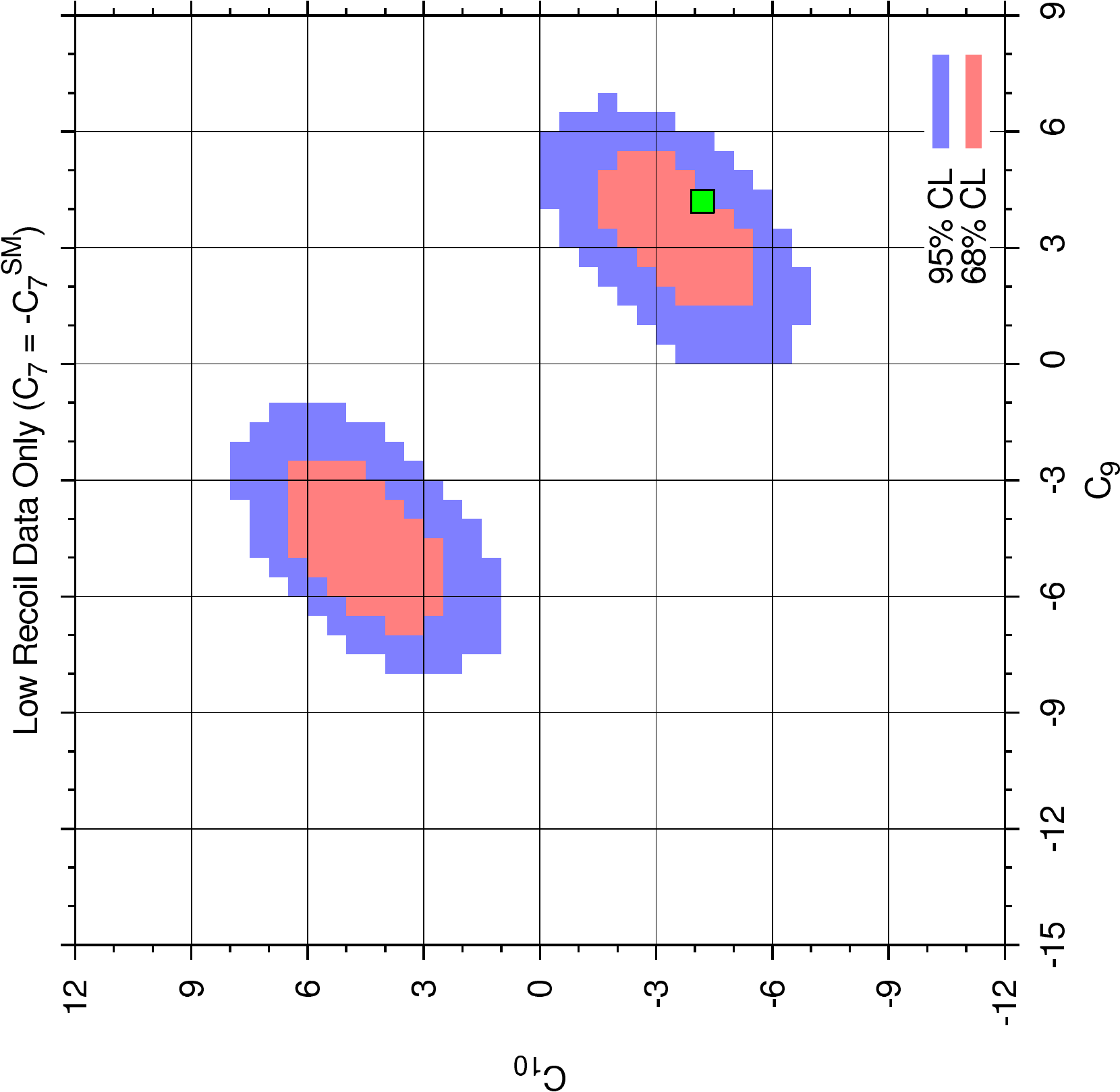}}
\end{center}
\caption{The constraints on $\wilson{9}$ and $\wilson{10}$ from 
$\bar{B}\to \bar{K}^*l^+l^-$ low recoil data \cite{:2009zv,CDF2010}  only
for $\wilson{7} = \wilson[SM]{7}$ (a)
  and  $\wilson{7} = -\wilson[SM]{7}$ (b) at $68\%$~CL (red areas)
  and $95\%$~CL (red and blue areas). The (green) square marks the SM value of ($\wilson{9},\wilson{10}$).}
\label{fig:constraintsHighQ2}
\end{figure}

\begin{figure}[b]
\begin{center}
  \includegraphics[angle=-90,width=.6\textwidth]{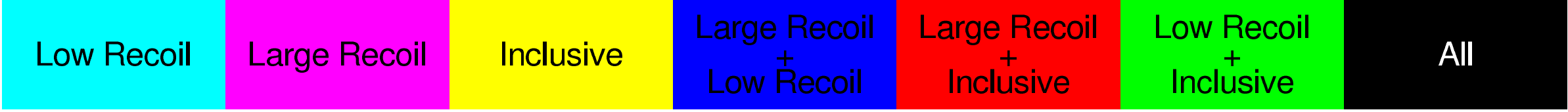}\\
  \subfigure[]{\includegraphics[angle=-90,width=.49\textwidth]{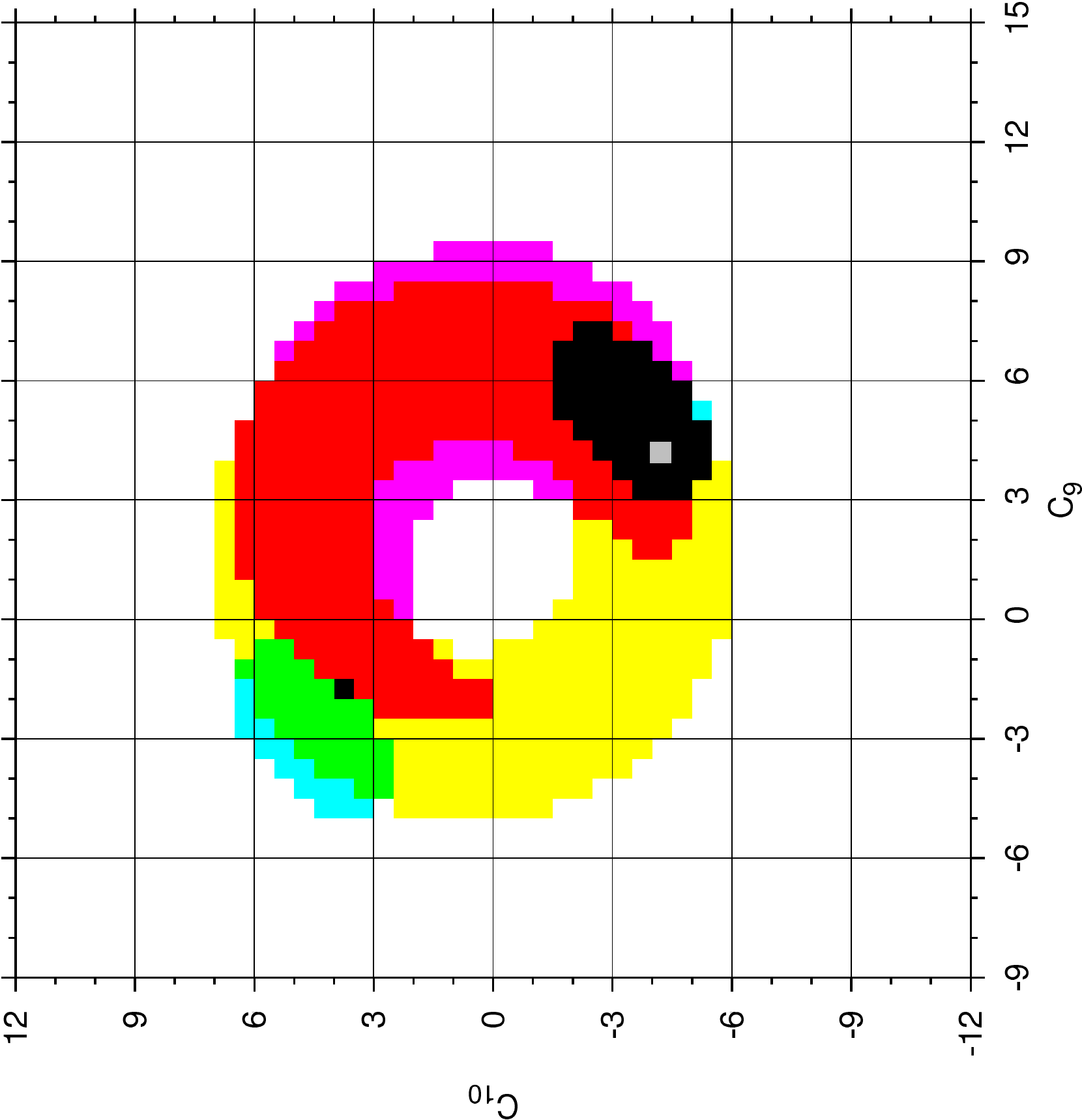}}
  \subfigure[]{\includegraphics[angle=-90,width=.49\textwidth]{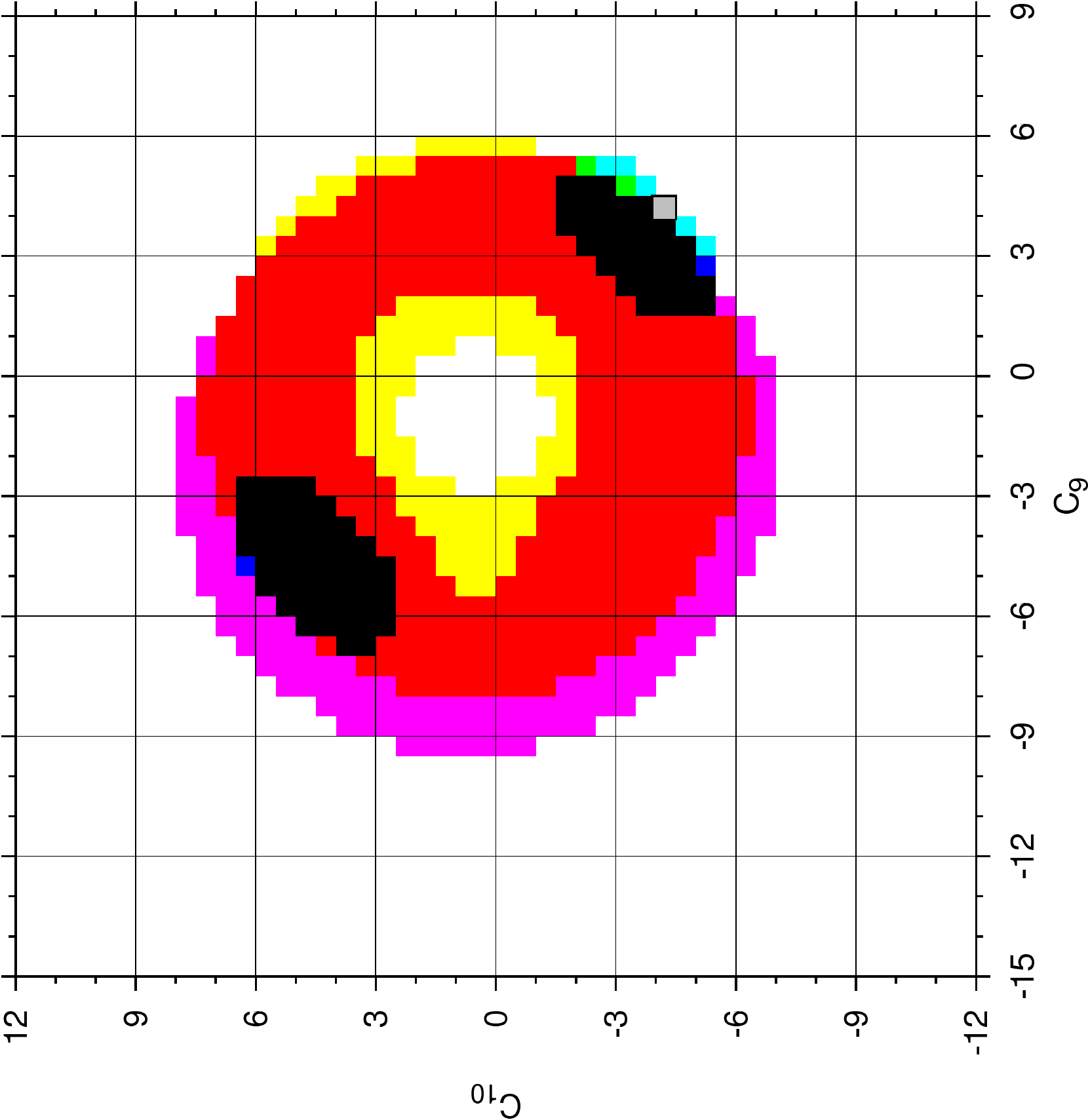}}
\end{center}
\caption{The individual $68\%$~CL constraints  on $\wilson{9}$ and $\wilson{10}$ 
  from  $\bar{B}\to \bar{K}^*l^+l^-$ at large and low recoil and $\bar{B}\to X_s l^+l^-$ 
  for $\wilson{7} = \wilson[SM]{7}$ (a) and  $\wilson{7} = -\wilson[SM]{7}$ (b)
  using Belle \cite{:2009zv,Iwasaki:2005sy}, BaBar \cite{Aubert:2004it} and CDF \cite{CDF2010} data.
  The (grey) square marks the SM value of ($\wilson{9},\wilson{10}$).
  See the color key at the top for the different constraints.}
\label{fig:constraintsIndividualOneSigma}
\end{figure}

\begin{figure}[b]
\begin{center}
  \subfigure[]{\includegraphics[angle=-90,width=.49\textwidth]{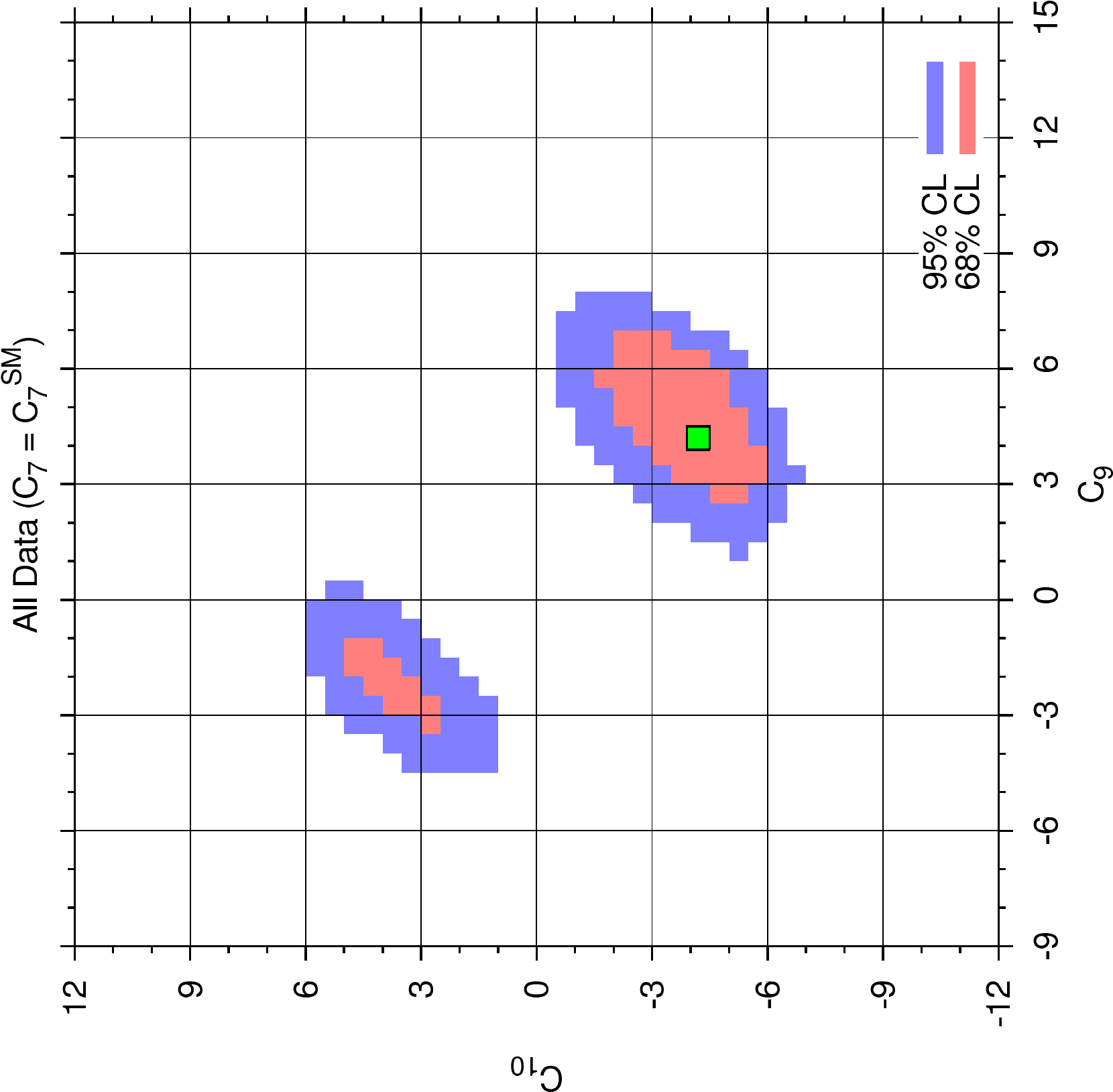}}
  \subfigure[]{\includegraphics[angle=-90,width=.49\textwidth]{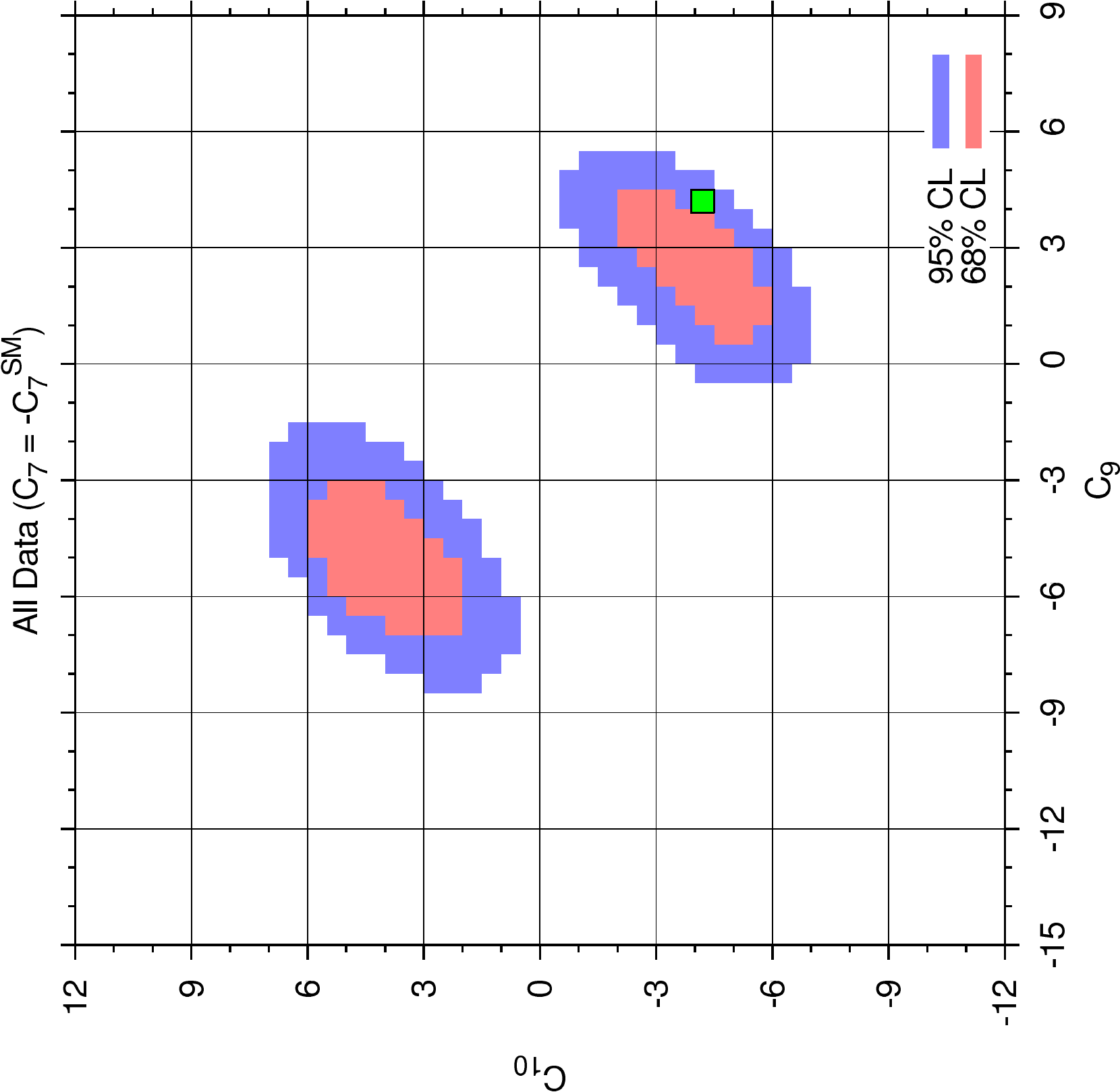}}
\end{center}
\caption{The global constraints on $\wilson{9}$ and $\wilson{10}$ from $\bar{B}\to \bar{K}^*l^+l^-$ and $\bar{B}\to X_s l^+l^-$ 
  for $\wilson{7} = \wilson[SM]{7}$ (a) and  $\wilson{7} = -\wilson[SM]{7}$ (b)
  using Belle \cite{:2009zv,Iwasaki:2005sy}, BaBar \cite{Aubert:2004it} and
  CDF \cite{CDF2010} data at $68\%$~CL (red area) and $95\%$~CL (red and blue areas).
  The (green) square marks the SM value of ($\wilson{9},\wilson{10}$).}
\label{fig:constraintsGlobal}
\end{figure}

To confront the available data with the SM we perform a parameter scan over
$-15 \leq \mathcal{C}_{9,10} \leq 15$ for $60\times 60$ points and check
the goodness-of-fit for each of the observables listed in \reftab{numericObservables}
in every point $(\mathcal{C}_9, \mathcal{C}_{10})$. We implement every observable analytically
with the single exception the $\bar B \to X_s \gamma$ branching ratio, for which we use the numerical
SM results given in \cite{Misiak:2006zs}. Contributions to the latter from physics beyond the SM
are implemented at leading order.
The integrated observables $\langle X\rangle_{q^2_{\rm min},q^2_{\rm max}}$ follow the
definition Eq. (\ref{eq:integratedObservables}) with the lower (upper)
integration boundary $q^2_{\rm min}$ ($q^2_{\rm max}$).
In particular we calculate
\begin{align}
  \chi_{i,\mathrm{E}}\(\{\wilson{j}\}\) & \equiv
  \begin{cases}
    \frac{|X_{i,\mathrm{T}} - X_{i,\mathrm{E}}| - \Delta^+_{i,\mathrm{T}}}
         {\sigma_{i,\mathrm{E}}}
    & X_{i,\mathrm{E}} \geq X_{i,\mathrm{T}} + \Delta^+_{i,\mathrm{T}}
\\
    \frac{|X_{i,\mathrm{T}} - X_{i,\mathrm{E}}| - \Delta^-_{i,\mathrm{T}}}
         {\sigma_{i,\mathrm{E}}}
    & X_{i,\mathrm{E}} \leq X_{i,\mathrm{T}} - \Delta^-_{i,\mathrm{T}}
\\
    0 & \text{otherwise}
  \end{cases}
  \label{eq:numericChi}
\end{align}
with the theoretical prediction of the $i$-th observable
$X_{i,\mathrm{T}} \equiv X_{i,\mathrm{T}}(\big\lbrace\wilson{j}\big\rbrace)$
and its upper (lower) uncertainty $\Delta^{+(-)}_{i,\mathrm{T}} =
\Delta^{+(-)}_{i,\mathrm{T}}(\big\lbrace\wilson{j}\big\rbrace)$ as
described in Section~\ref{sec:smstatus}.
The experimental result from experiment E for the
  $i$-th observable is denoted by $X_{i,{\rm E}}$ and its error
  $\sigma_{i,E}$ is obtained by adding linearly the statistical and
  systematic errors and subsequent symmetrization.
{}From here we calculate the likelihood $\mathcal{L}$ as
\begin{align}
  \mathcal{L}(\{\wilson{j}\}) & =
    \exp \[ -\frac{1}{2}  \sum_{i,\mathrm{E}} \chi^2_{i,\mathrm{E}}(\{\wilson{j}\}) \]
  \label{eq:numericLikelihood}.
\end{align}

These scans allow us to constrain the values of the coefficients $\wilson{9}$ and $\wilson{10}$
under the assumption that they are real-valued, \emph{i.e.}, there is no
CP violation beyond the SM, and that $\wilson{1\dots6,8}$ take on their respective SM values.
The $\mathcal{B}(\bar B\to X_s\gamma)$ data constrain the magnitude of  $\wilson{7}$ strongly to
a narrow range of values around $ |\wilson[SM]{7}|$, however without
determining the sign of $\wilson{7}$. For this reason, we present in the following our scans for
$\wilson{7} = \pm \wilson[SM]{7}$.

In \reffig{constraintsLowQ2+Inclusive} we show the constraints in the 
$\wilson{9}-\wilson{10}$ plane from
$\bar{B}\to \bar{K}^*l^+l^-$ decays at large recoil and $\bar{B}\to X_sl^+l^-$ data, without
use of the low recoil information. On the other hand,
taking into account  the $\bar{B}\to\bar{K}^*l^+l^-$ data at low recoil only,
we arrive at the constraints given in \reffig{constraintsHighQ2}. We see that the latter low recoil constraints  are presently much more powerful than the others.
An important ingredient for this are the $A_{\rm FB}$ measurements at low recoil 
constraining $A_{\rm FB} \propto \Re{\wilson{9}\wilson[*]{10}}$ to be SM-like,
the benefits of which have already been pointed out in \cite{Bobeth:2008ij}.
The individual constraints, overlaid on top of each other, are given at $68\%$~CL 
 in \reffig{constraintsIndividualOneSigma}. The data are consistent with each other.

The global constraints, obtained after summing over the 
$\chi^2$-values of all aforementioned data, are shown in \reffig{constraintsGlobal}.
Two disjoint solutions are favored, 
around $(\wilson[SM]{9},\wilson[SM]{10})$ or in the vicinity
of $(-\wilson[SM]{9},-\wilson[SM]{10})$. There appears to be space for order one deviations from either solution, regardless of the sign
of $\wilson{7}$. Note that  the flipped-sign solution around $(-\wilson[SM]{9},-\wilson[SM]{10})$
for $\wilson{7} = \wilson[SM]{7}$ is disfavored, see  \reffig{constraintsIndividualOneSigma}.
Varying $\wilson{7}$ between -0.5 and +0.5
and imposing the $\bar B\to X_s\gamma$ constraint leads to barely noticable larger contours
in the $\wilson{9}-\wilson{10}$ plane than the ones in \reffig{constraintsGlobal} a (for $\wilson{7}<0$)
and \reffig{constraintsGlobal}  b (for $\wilson{7}>0$),
and are not shown.

We find that at $2\sigma$ the allowed values of $\wilson{10}$ are within
$0.5 \leq |\wilson{10}| \leq 8$. This gives branching ratios for 
$\bar{B}_s \to \mu^+ \mu^-$ decays enhanced or lowered with respect to the SM one,
within  the interval $[2 \times 10^{-11}, 1.3\times 10^{-8}]$. This is consistent with the
current upper limit  on this mode, ${\cal{B}}(\bar{B}_s \to \mu^+ \mu^-) < 3.6 \times 10^{-8}$ ($95\%$~CL) \cite{Barberio:2008fa}.
Similarly, the values of the transversity observables $\langle H_T^{(2,3)} \rangle$ integrated over
the low recoil region, \refeq{loreco-range}, are  within
the ranges  $-1.0$ and  $+0.2$.

As the  experimental precision improves over time, especially with the LHC$b$ data at the horizon, there will be opportunities to resolve the 4-fold ambiguity of the current solutions presented in \reffig{constraintsGlobal}.
Firstly, knowing whether $A_{\rm FB}$ has a zero for low $q^2$ as in the SM or not,
fixes the sign of ${\rm Re} \{\wilson{7} {\cal{C}}_{10}^{*} \}$,
thereby eliminating two of the four possible solutions.
Alternatively, the sign of the interference term $\Re{\wilson[*]{7}\wilson{9}}$ in
$\mathcal{B}(\bar B\to X_sl^+l^-)$ can be extracted from precision measurements.
In the SM, this term decreases
the branching ratio. These two effects are correlated within our framework, {\it i.e.}, 
the existence of an $A_{\rm FB}$ zero crossing implies a destructive
 interference term in the branching ratio and vice versa.

At this point,  there would still be two possible solutions left. Assuming, for instance, a confirmation of the $A_{\rm FB}$ zero,  these solutions are $\wilson{7,9,10}$ having
SM-like signs, or $\wilson{7,9,10}$ having opposite signs with respect
to their SM values. This last ambiguity can be resolved with precision measurements at the
level where one becomes sensitive to the (known) difference between the 
Wilson coefficients $\wilson{i}$
and the effective ones $\wilson[eff]{i}$.
Then,  the additional contribution breaks the symmetry in the observables under sign reflection.
Since the contribution of $\wilson{7}$ to the decay amplitudes is small at large $q^2$,  promising  observables
to resolve the final sign issue are  those at low dilepton masses.

\begin{table}
\begin{tabular}{l|lr|lr}
\hline \hline
\multicolumn{1}{c|}{Observable} & \multicolumn{2}{c|}{SM Prediction} &
\multicolumn{2}{c}{Measurement}\\
\hline \hline
  $\int^{6.0}_{1.0} \dd q^2 \dd \mathcal{B}(\bar B\to X_s l^+l^-)/\dd q^2$
& $\(1.55 \pm 0.11\) \times 10^{-6}$ & \cite{Bobeth:2003at}
& $\(1.49^{+0.92}_{-0.83}\) \times 10^{-6}$ & \cite{Iwasaki:2005sy}\\
& & & $\(1.8 \pm 1.2\) \times 10^{-6}$ & \cite{Aubert:2004it}\\
\hline
  $\mathcal{B}(\bar B\to X_s \gamma)_{E_\gamma > 1.6\GeV}$
& $\(3.15 \pm 0.23\) \times 10^{-4}$ & \cite{Misiak:2006zs}
& $\(3.55 \pm 0.33\) \times 10^{-4}$ & \cite{Barberio:2008fa}\\
\hline
  $\langle  \mathcal{B}(\bar{B}\to\bar{K}^*l^+l^-)\rangle_{1.0,6.0}$
& $\(2.60^{+1.82}_{-1.34}\) \times 10^{-7}$ & \cite{Bobeth:2008ij}
& $\(1.49^{+0.57}_{-0.52}\) \times 10^{-7}$ & \cite{:2009zv}\\
& & & $\(1.60^{+0.68}_{-0.68}\) \times 10^{-7}$ & \cite{CDF2010}\\
\hline
  $\langle A_{\rm FB}(\bar{B}\to\bar{K}^*l^+l^-)\rangle_{1.0,6.0}$
& $+0.05^{+0.04}_{-0.03}$ & \cite{Bobeth:2008ij}
& $-0.26^{+0.37}_{-0.34}$ & \cite{:2009zv}\\
& & & $-0.43^{+0.43}_{-0.42}$ & \cite{CDF2010}\\
\hline
  $\langle F_{\rm L}(\bar{B}\to\bar{K}^*l^+l^-)\rangle_{1.0,6.0}$
& $0.73^{+0.13}_{-0.23}$ & \cite{Bobeth:2008ij}
& $0.67^{+0.28}_{-0.28}$ & \cite{:2009zv}\\
& & & $0.50^{+0.30}_{-0.33}$ & \cite{CDF2010}\\
\hline
  $\langle  \mathcal{B}(\bar{B}\to\bar{K}^*l^+l^-)\rangle_{14.18,16.00}$
& $\(1.32^{+0.43}_{-0.36}\) \times 10^{-7}$ & \cite{Bobeth:2008ij}
& $\(1.05^{+0.37}_{-0.34}\) \times 10^{-7}$ & \cite{:2009zv}\\
& & & $\(1.51^{+0.49}_{-0.49}\) \times 10^{-7}$ & \cite{CDF2010}\\
\hline
  $\langle A_{\rm FB}(\bar{B}\to\bar{K}^* l^+ l^-)\rangle_{14.18,16.00}$
& $-0.44^{+0.07}_{-0.07}$ &
& $-0.70^{+0.32}_{-0.26}$ & \cite{:2009zv}\\
&&  & $-0.42^{+0.25}_{-0.25}$ & \cite{CDF2010}\\
\hline
  $\langle  \mathcal{B}(\bar{B}\to\bar{K}^*l^+l^-)\rangle_{16.00,19.21}$
& $\(1.54^{+0.48}_{-0.42}\) \times 10^{-7}$ & \cite{Bobeth:2008ij}
& $\(2.04^{+0.43}_{-0.40}\) \times 10^{-7}$ & \cite{:2009zv}\\
& & & $\(1.35^{+0.49}_{-0.49}\) \times 10^{-7}$ & \cite{CDF2010}\\
\hline
  $\langle A_{\rm FB}(\bar{B}\to\bar{K}^* l^+ l^-)\rangle_{16.00,19.21}$
& $-0.38^{+0.07}_{-0.07}$ &
& $-0.66^{+0.20}_{-0.15}$ & \cite{:2009zv}\\
&&  & $-0.70^{+0.35}_{-0.26}$ & \cite{CDF2010}\\
\hline
\hline
\end{tabular}
\caption{The observables used to constrain $\wilson{7,9,10}$. The experimental data
  for $A_{\rm FB}$ have their sign flipped to match the conventions used in this
  work. For the notation of the observables, see text. }
  \label{tab:numericObservables}
\end{table}

%
%--------+---------+---------+---------+---------+---------+---------+---------+
\section{Conclusions \label{sec:con}}

Discrepancies between $b$ physics predictions and measurements can be caused
by new physics beyond the SM or by an insufficiently accounted for background from strong interaction bound state effects.  Due to the decays  simple transversality structure at low recoil, these QCD and electroweak effects can be disentangled in 
$\bar B \to \bar K^* l^+ l^-$ angular studies.

In fact, to leading order in the power corrections with subleading terms being further 
suppressed, all contributing transversality amplitudes 
exhibit the same dependence on the short distance electroweak physics, which moreover
factorizes from the hadronic matrix elements.
This in turn allows to define new observables, $H_{T}^{(1,2,3)}$, see \refeqs{def:HT1}{def:HT3}, which do not depend on the form factors at low recoil
 and cleanly test the SM. Other observables, which do not depend on the Wilson coefficients
at low recoil, such as $F_{\rm L}$, $A_T^{(2,3)}$ and the newly constructed ones in 
\refeq{ff-ratio-test},  probe certain $B \to K^*$ form factors combinations.
Measurements of the latter provide input to form factor parametrizations along the lines of
 \cite{Bharucha:2010im},  which could be compared to (future) lattice results.

Exploiting data we find that the constraints from the low recoil region add significant new information, while being consistent with the large recoil and inclusive decays data, and the
SM. Large deviations from the SM are, however, allowed as well due to the current experimental uncertainties.
Our findings are summarized in Figs.~\ref{fig:observables-in-SM} and \ref{fig:constraintsGlobal}.
Improved measurements of the forward-backward asymmetry  or precision data on
the inclusive $\bar B \to X_s l^+ l^-$ branching ratio can resolve the present ambiguities in the 
best-fit solution.

Since  the decay $\bar{B}_s \to \phi \mu^+ \mu^-$ has
been seen \cite{CDF2010}, it becomes relevant in the near future as
well.  The low recoil framework and our analysis applies to $B_s$ decays with the obvious
replacements of masses and hadronic input. 

To conclude, we obtained from the existing data on $\bar B \to \bar K^* l^+ l^-$ decays at low recoil new and most powerful constraints. The proposed angular studies offer great opportunities,  
both in terms of consistency checks and precision, to explore further the borders of the SM.

%--------+---------+---------+---------+---------+---------+---------+---------+
\acknowledgments
 
 We thank Hideki Miyake for  useful communication on CDFs $\bar B \to \bar K^* \mu^+ \mu^-$ analysis and Ben Grinstein  for useful comments on the manuscript. We are grateful
 to Dan Pirjol for comments and numerical checks.
This work is supported in part by the Bundesministerium f\"ur Bildung und
Forschung.

%--------+---------+---------+---------+---------+---------+---------+---------+
%
% Appendix
%
%--------+---------+---------+---------+---------+---------+---------+---------+
\appendix

%
%
%--------+---------+---------+---------+---------+---------+---------+---------+
\section{The Angular Coefficients  \label{sec:JifromAi}}

Here, the coefficients $J_i^{(a)}$ in the angular distribution
\refeq{decaydistribution} are given in terms of the transversity amplitudes
$A_{\perp,\parallel,0,t}$ \cite{Kruger:2005ep}. Terms with finite lepton masses,
which are of relevance at low $q^2$, have been kept.
\begin{align}\label{I}
  J_1^s & = \frac{3}{4}  \{ \frac{(2+\beta_l^2)}{4} \left[|\apeL|^2 + |\apaL|^2 + (L\to R) \right] + \frac{4 m_l^2}{q^2} \re\left(\apeL^{}\apeR^* + \apaL^{}\apaR^*\right)  \}, 
\\
J_1^c & = \frac{3}{4} \{ |\azeL|^2 +|\azeR|^2  + \frac{4m_l^2}{q^2}  \left[|A_t|^2 + 2\re(\azeL^{}\azeR^*) \right] \},
\\
  J_2^s & = \frac{3 \beta_l^2}{16}\bigg[ |\apeL|^2+ |\apaL|^2 + (L\to R)\bigg],
\\
  J_2^c & = - \frac{3\beta_l^2}{4}\bigg[|\azeL|^2 + (L\to R)\bigg],
\\
  J_3 & = \frac{3}{8}\beta_l^2\bigg[ |\apeL|^2 - |\apaL|^2  + (L\to R)\bigg],
\\
  J_4 & = \frac{3}{4\sqrt{2}}\beta_l^2\bigg[\re (\azeL^{}\apaL^*) + (L\to R)\bigg],
\\
  J_5 & = \frac{3\sqrt{2}}{4}\beta_l\bigg[\re(\azeL^{}\apeL^*) - (L\to R)\bigg],
\\
  J_6 & = \frac{3}{2}\beta_l\bigg[\re (\apaL^{}\apeL^*) - (L\to R)\bigg],
\\
  J_7 & = \frac{3\sqrt{2}}{4} \beta_l \bigg[\im (\azeL^{}\apaL^*) - (L\to R)\bigg],
\\
  J_8 & = \frac{3}{4\sqrt{2}}\beta_l^2\bigg[\im(\azeL^{}\apeL^*) + (L\to R)\bigg],
\\
  J_9 & = \frac{3}{4}\beta_l^2\bigg[\im (\apaL^{*}\apeL) + (L\to R)\bigg],
\end{align}
where
\begin{equation}
  \beta_l = \sqrt{1 - \frac{4 m_l^2}{q^2}} .
\end{equation}
The transversity amplitudes at low recoil are given in Section \ref{sec:loreco}.
The ones at large recoil can be seen in \cite{Bobeth:2008ij}.

%
%
%--------+---------+---------+---------+---------+---------+---------+---------+
\section{The Low Recoil Transversity Observables 
  \label{sec:Ui} }
  
It is useful to  introduce the ($q^2$-dependent) quantities 
\begin{align}
  U_1 & = |A_0^L|^2 + |A_0^R|^2, &
  U_4 & = {\rm Re} (A_0^L A_\parallel^{L*} + A_0^{R*} A_\parallel^R), &
  U_7 & = {\rm Im} (A_0^L A_\parallel^{L*} + A_0^{R*} A_\parallel^R), \\
  U_2 & = |A_\perp^L|^2 + |A_\perp^R|^2,& 
  U_5 & = {\rm Re} (A_0^L A_\perp^{L*} - A_0^{R*} A_\perp^R), &
  U_8 & = {\rm Im} (A_0^L A_\perp^{L*} - A_0^{R*} A_\perp^R),\\
  U_3 & = |A_\parallel^L|^2 + |A_\parallel^R|^2, &
  U_6 & = {\rm Re} (A_\parallel^L A_\perp^{L*} - A_\parallel^{R*} A_\perp^R), &
  U_9 & = {\rm Im} (A_\parallel^L A_\perp^{L*} - A_\parallel^{R*} A_\perp^R),
\end{align}
which are invariant under the transformations \cite{Egede:2008uy}
\begin{align}
  \nonumber
  A_i^{L,R} & \mapsto e^{i\phi_{L,R}} A_i^{L,R}, &
\\
  \label{eq:d4G:sym:trafo}
  A_{0,\parallel}^{L,R} & \mapsto \cos\theta A_{0,\parallel}^{L,R} \mp 
      \sin\theta A_{0,\parallel}^{R,L*}, &
  A_\perp^{L,R} & \mapsto \cos\theta A_\perp^{L,R} \pm \sin\theta A_\perp^{R,L*}.
\end{align}
In the limit
$m_l\to 0$ the decay rate ${\rm d}^4\Gamma$ becomes invariant under the transformations
\refeq{d4G:sym:trafo} as well, reducing the number of 11 independent $J_i^{(a)}$ to 9
\cite{Egede:2008uy}.

In terms of the $J_i$, the $U_i$ read
\begin{align}
  U_1 & = -\frac{4}{3 \beta_l^2} J_2^c, &
  U_2 & = \frac{4}{3 \beta_l^2} \left[2 J_2^s + J_3\right], &
  U_3 & = \frac{4}{3 \beta_l^2} \left[2 J_2^s -  J_3\right], 
\nonumber \\
  U_4 & = \frac{\sqrt{32}}{3 \beta_l^2} J_4, &
  U_5 & = \frac{\sqrt{8}}{3 \beta_l} J_5, &
  U_6 & = \frac{2}{3 \beta_l} J_6,
\\
  U_7 & = \frac{\sqrt{8}}{3 \beta_l} J_7, &
  U_8 & = \frac{\sqrt{32}}{3 \beta_l^2} J_8, &
  U_9 & = -\frac{4}{3 \beta_l^2} J_9. 
\nonumber
\end{align}
Whereas we refer to the $J_i^{(a)}$ as observables which can be extracted
from the angular analysis, the $U_i$  greatly simplify the discussion of the 
form factor related uncertainties.

At high $q^2$, where we can set $m_l$ to zero safely, the $U_i$ factorize into the 
short distance $\rho_{1,2}$ and three independent form factor coefficients
$f_i$, $i=0,\perp,\parallel$ given in \refsec{transA} as
\begin{align}
  U_1 & = 2 \rho_1 f_0^2, &
  U_2 & = 2 \rho_1 f_\perp^2, &
  U_3 & = 2 \rho_1 f_\parallel^2,
\nonumber \\ 
  U_4 & = 2 \rho_1 f_0 f_\parallel, &
  U_5 & = 4 \rho_2 f_0 f_\perp, &
  U_6 & = 4 \rho_2 f_\parallel f_\perp, &
  U_7 & = U_8 = U_9 = 0.
  \label{eq:U:HQET:limit}
\end{align}
The coefficients $J_7,J_8,J_9$ and likewise $U_7, U_8,U_9$ vanish, since they are proportional to $\im
(A_i^X A_j^{X *})$, $i,j=0,\parallel, \perp$ and by means of the identical short distance 
dependence of the transversity amplitudes Eqs.~(\ref{eq:Aperp})-(\ref{eq:A0}).

The simple and factorizable  structure of \refeq{U:HQET:limit} allows to test at the same time the SM and  the
hadronic input used. Firstly, one
can construct three independent low recoil transversity observables free of form
factors in the HQET symmetry limit, which we define as
\begin{align}
  H_T^{(1)} & = \frac{U_4}{\sqrt{U_1 \cdot U_3}} ,
  \label{eq:def:H1}
\\
  H_T^{(2)} & = \frac{U_5}{\sqrt{U_1 \cdot U_2}},
  \label{eq:def:H2}
\\
  H_T^{(3)} & =\frac{U_6}{\sqrt{U_2 \cdot U_3}},
  \label{eq:def:H3}
\end{align} 
see \refeq{Hi:HQET} and Section \ref{sec:observables} for their 
explicit expressions in terms of the $J_i$ and the transversity amplitudes.
Secondly, ratios of the form factors $f_j/f_k$ can be tested
independently of the short distance couplings $\rho_i$ using the observables
\begin{align} \label{eq:ff-ratio-test}
  \frac{f_0}{f_\parallel} & = \sqrt{\frac{U_1}{U_3}} = \frac{U_1}{U_4} 
    = \frac{U_4}{U_3} = \frac{U_5}{U_6}, &
  \frac{f_0}{f_\perp} & = \sqrt{\frac{U_1}{U_2}},  &
  \frac{f_\perp}{f_\parallel} & = \sqrt{\frac{U_2}{U_3}} 
    = \frac{\sqrt{U_1 U_2}}{U_4}.
\end{align}

%
%
%--------+---------+---------+---------+---------+---------+---------+---------+
\section{The Form Factors}
\label{sec:ChoiceFF}

The hadronic matrix elements of a $B$ meson with 4-momentum $p$ decaying into a
vector meson can be parametrized as \cite{Ball:2004rg}:
\begin{align}
    \bra{V(k, \epsilon)} \bar{q} \gamma_\mu b \ket{B(p)}\label{eq:ffbasisBZfirst}
        & = \frac{2 V(q^2)}{m_B + m_V} \varepsilon_{\mu\rho\sigma\tau} \epsilon^{*\rho} p^\sigma k^\tau ,\\
    \bra{V(k, \epsilon)} \bar{q} \gamma_\mu \gamma_5 b \ket{B(p)}
        & = i\epsilon^{*\rho}
            \[2 m_V A_0(q^2) \frac{q_\mu q_\rho}{q^2} + (m_B + m_V) A_1(q^2)\(g_{\mu\rho} - \frac{q_\mu q_\rho}{q^2}\)\right. \nonumber \\
     & - \left.A_2(q^2) \frac{q_\rho}{m_B + m_V}\((p+k)_\mu - \frac{m_B^2 - m_V^2}{q^2} (p-k)_\mu\)\] ,\\
    \bra{V(k, \epsilon)} \bar{q} i \sigma_{\mu\nu} q^\nu b\ket{B(p)}
        & = -2T_1(q^2) \varepsilon_{\mu\rho\sigma\tau} \epsilon^{*\rho} p^\sigma k^\tau, \\
    \bra{V(k, \epsilon)} \bar{q} i \sigma_{\mu\nu} \gamma_5 q^\nu b\ket{B(p)}
        & = iT_2(q^2) \(\epsilon^*_\mu (m_B^2 - m_V^2) - (\epsilon^* \cdot q) (p+k)_\mu\)  \nonumber
        \\
         & + iT_3(q^2) \(\epsilon^* \cdot q\)\(q_\mu - \frac{q^2}{m_B^2 - m_V^2}(p+k)_\mu\) ,\label{eq:ffbasisBZlast}
\end{align}
where $m_V,k$ and $ \epsilon$ denote the mass, 4-momentum and the polarization
vector of the vector meson, respectively. The seven form factors $V,A_{0,1,2}$
and $T_{1,2,3}$ are functions of the momentum transfer $q^2$, and $q=p-k$. Note
that by parity-invariance $ \bra{V(k, \epsilon)} \bar{q} b \ket{B(p)}=0$.

LCSR provide the form factors at large recoil, $q^2 \lesssim 14 \, \mbox{GeV}^2$
\cite{Ball:2004rg}. There, the outcome of the LCSR calculation is fitted to a
physical $q^2$ dependence, of pole or dipole structure. It is conceivable that
the form factor parametrization obtained in this way are valid at low recoil as
well.

\begin{figure}
    \centering
    \includegraphics[angle=-90,width=.7\textwidth]{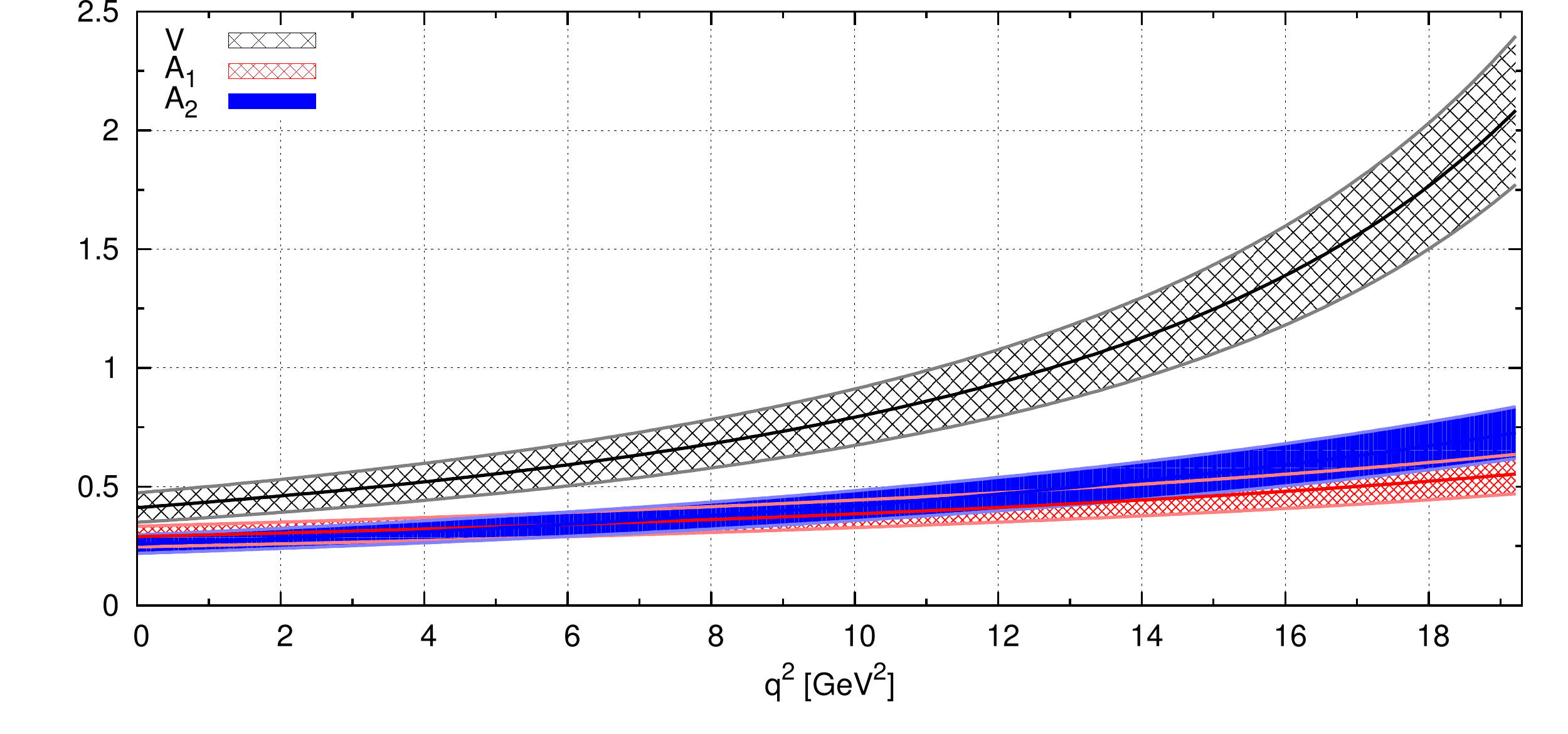}
    \caption{The $B \to K^*$ form factors $V,A_1$ and $A_2$ from \cite{Ball:2004rg}.}
\label{fig:VA12}
\end{figure}

For completeness, we give here the parametrization of the form factors $V,
A_{1,2}$ from \cite{Ball:2004rg}, which we use at both low and large recoil.
\begin{align}
    V(q^2)   & = \frac{r_1}{1 - q^2 / m_R^2} + \frac{r_2}{1 - q^2 / m_{\rm fit}^2},\\
    A_1(q^2) & = \frac{r_2}{1 - q^2 / m_{\rm fit}^2},\\
    A_2(q^2) & = \frac{r_1}{1 - q^2 / m_{\rm fit}^2} + \frac{r_2}{(1 - q^2/m_{\rm fit}^2)^2},
\end{align}
where the fit parameters $r_{1,2}, m_R^2$ and $m_{\rm fit}^2$ are given in
\reftab{ff-parameters}.  The resulting form factors are shown in \reffig{VA12}.
For the uncertainty we use 15 \% as follows from the LCSR calculation.

\begin{table}
    \begin{tabular}{c|cccc}
    \hline
    \hline
          & $r_1$    & $r_2$    & $m_R^2~[\GeV^2]$ & $m_{\rm fit}^2~[\GeV^2]$\\
    \hline
    $V$   & $0.923$  & $-0.511$ & $5.32^2$         & $49.40$\\
    $A_1$ & --       & $0.290$  & --               & $40.38$\\
    $A_2$ & $-0.084$ & $0.343$  & --               & $52.00$\\
    \hline
    \hline
    \end{tabular}
    \caption{The parameters of the form factors $V, A_{1,2}$. \label{tab:ff-parameters}}
\end{table}

% T_{1,2}
\begin{figure}
    \centering
    \subfigure[]{\includegraphics[angle=-90,width=.7\textwidth]{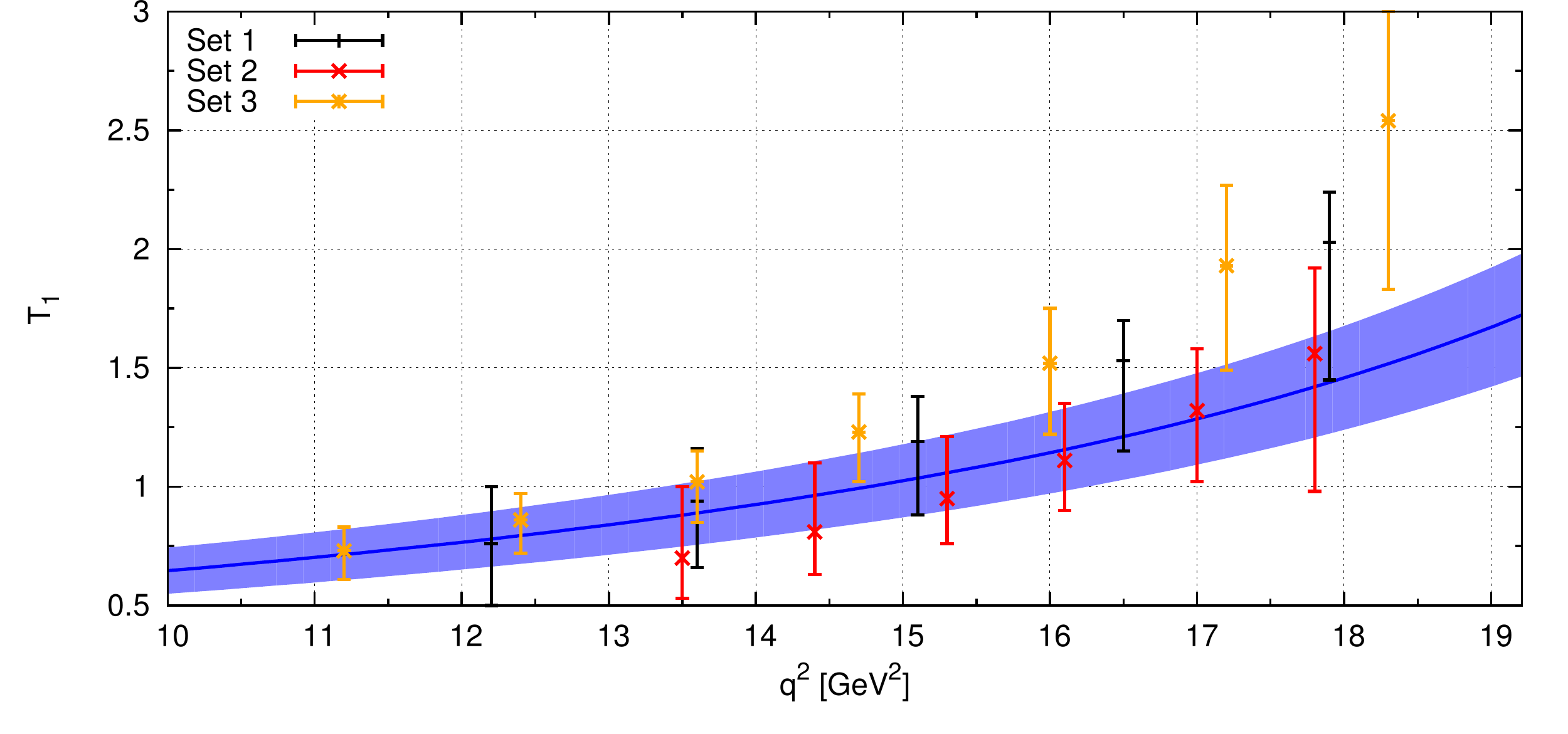}}\\
    \subfigure[]{\includegraphics[angle=-90,width=.7\textwidth]{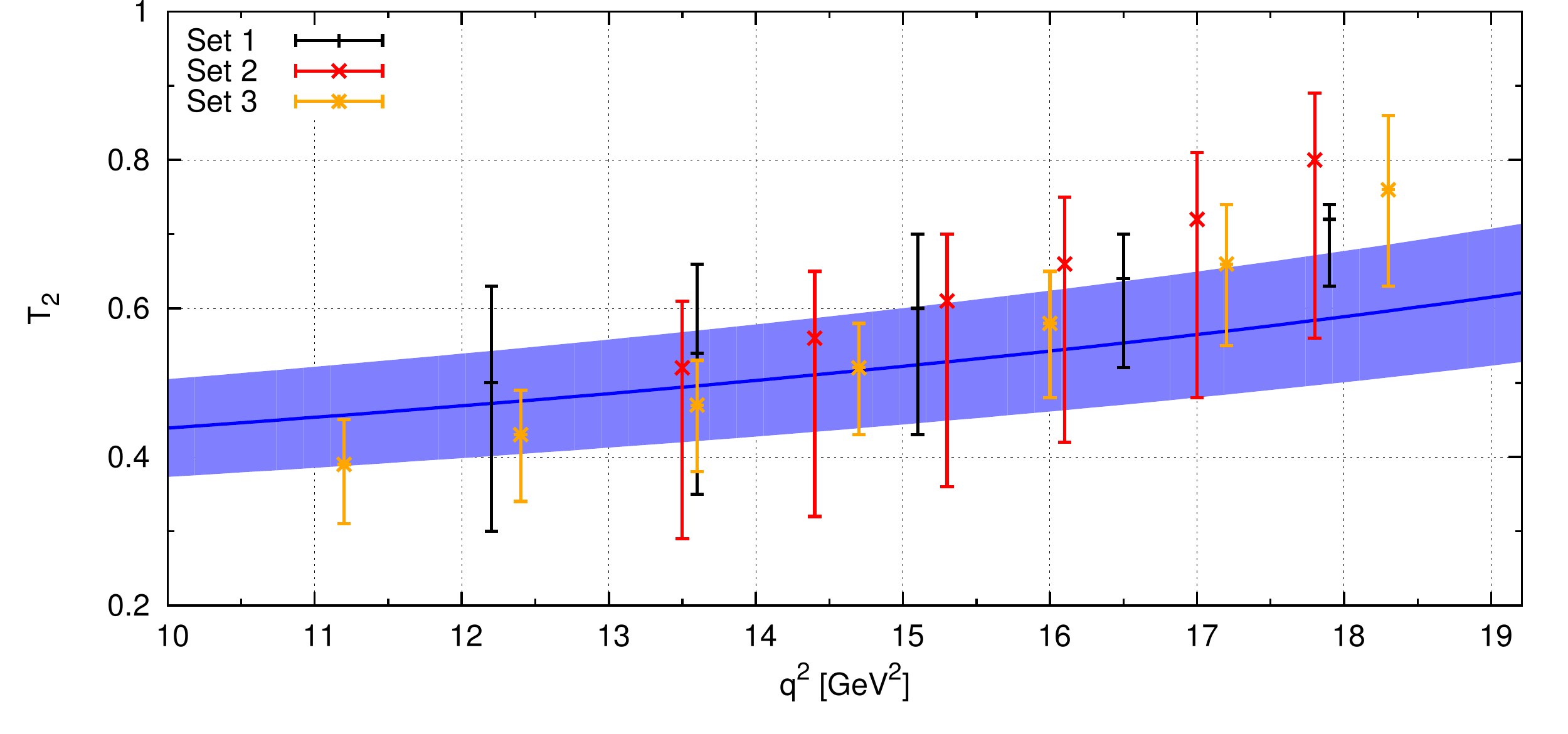}}
    \caption{The form factors $T_1$ (a) and $T_2$ (b) for $B \to K^*$
             transitions from \cite{Ball:2004rg} (blue bands) and lattice QCD
             results (3 data sets)
             \cite{Becirevic:2006nm}.\label{fig:t1_and_t2} }
             \end{figure}

In \reffig{t1_and_t2} we compare the LCSR fit against the lattice results, which
exist for $T_{1,2}$ \cite{Becirevic:2006nm}. The agreement is reasonable, given
the substantial uncertainties. There is consistency as well with the preliminary unquenched
findings  of Ref.~\cite{Liu:2009dj}, which are not shown.

\begin{figure}
    \centering
    \includegraphics[angle=-90,width=.7\textwidth]{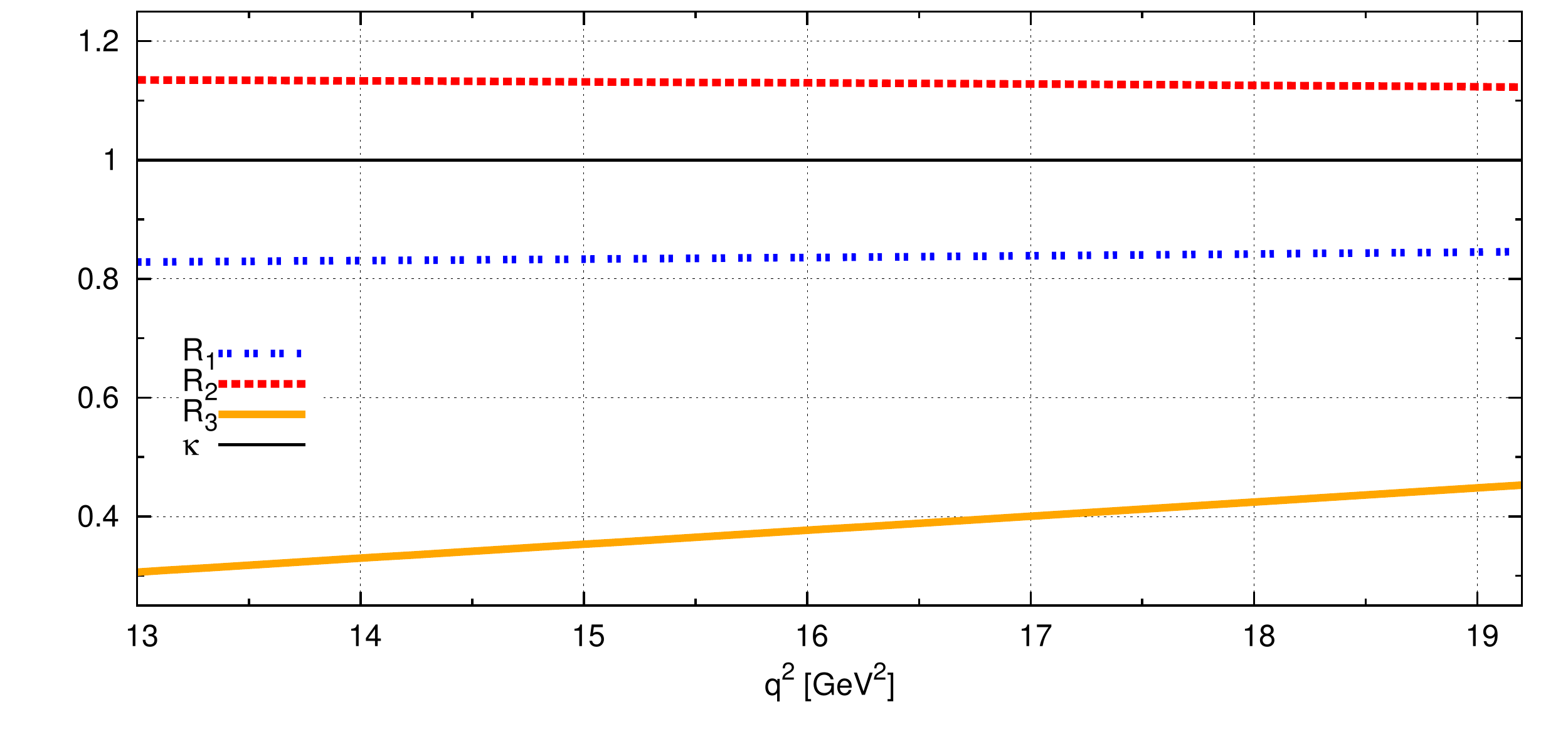}\\
    \caption{Comparison of the extrapolated LCSR form factors from
             \cite{Ball:2004rg} to the improved Isgur-Wise relations
             \refeq{ffrelations}. Shown is $R_1$ (blue dotted line), $R_2$ (red
             dashed line) and $R_3$ (golden solid line) as given in
             \refeq{quality-isgur-wise} and $\kappa=1 +{\cal{O}}(\alpha_s^2)$      for
             $\mu=m_b(m_b)$
             (black thick line).} \label{fig:isgur-wise-relations} 
             \end{figure}

How well do the LCSR form factors from \cite{Ball:2004rg} satisfy the low recoil
form factor relations \refeq{ffrelations}?  In \reffig{isgur-wise-relations} we
show the ratios
\begin{equation} \label{eq:quality-isgur-wise}
    R_1 =\frac{T_1(q^2)}{V(q^2)},\qquad R_2 = \frac{T_2(q^2)}{A_1(q^2)},\qquad R_3 =\frac{q^2}{m_B^2}\frac{T_3(q^2)}{A_2(q^2)},
\end{equation}
which in the symmetry limit should all equal $\kappa$, which is also shown.
Note, that in the large energy limit $E_{K^*} \gg \Lambda$ 
the form factors obey to lowest order in the strong coupling very similar relations $R_{1,2}=1 + {\cal{O}}(m_{K^*}/m_B)$ and $T_3/A_2=1 + {\cal{O}}(m_{K^*}/m_B)$
 \cite{Burdman:2000ku,Charles:1998dr}.
We learn that the improved Isgur-Wise relations work reasonably well for 
the extrapolated LCSR form factors with the exception of the one for $T_3$.
The agreement improves here somewhat if the factor $q^2/m_B^2$ is replaced by 
one, its leading term in the heavy quark expansion.

% Isgur-Wise relations

For the low $q^2$ form factors we employ a factorization scheme within QCDF 
where the $\xi_{\perp, \parallel}$ are related to the $V, A_{1,2}$ 
as \cite{Beneke:2004dp}
\begin{align}
  \label{eq:xi:def}
  \xi_\perp & = \frac{m_B}{m_B + m_{K^*}} V, &
  \xi_\parallel & = \frac{m_B + m_{K^*}}{2 E_{K^*}} A_1 -
                    \frac{m_B - m_{K^*}}{m_B} A_2.
\end{align}
%--------+---------+---------+---------+---------+---------+---------+---------+
%
% Bibliography
%
%--------+---------+---------+---------+---------+---------+---------+---------+

\end{document}